\documentclass[12pt,preprint]{aastex}

\newcommand{\ud}{\mathrm{d}}
\newcommand{\mm}{$\mu$m}

\newcommand{\app}{Astroparticle Physics}
\newcommand{\ICRC}{International Cosmic Ray Conference}

\begin{document}


\title{Upper Limits on the Extragalactic Background Light from the Very High Energy Gamma-Ray Spectra of Blazars}


\author{M. Schroedter\altaffilmark{1}}
\affil{Whipple Observatory, Harvard-Smithsonian Center for Astrophysics, P.O. Box 6369, Amado, AZ 85645-0097}
\email{mschroed@cfa.harvard.edu}
\affil{present address: Dept. of Physics and Astronomy, Iowa State University,
    Ames, IA 50011-3160}



\begin{abstract}
The direct measurement of the extragalactic background light (EBL) is difficult at optical to infrared wavelengths because of the strong foreground radiation originating in the Solar System. Very high energy (VHE, E$>$100 GeV) gamma rays interact with EBL photons of these wavelengths through pair production. In this work, the available VHE spectra from six blazars are used to place upper limits on the EBL. These blazars have been detected over a range of redshifts and a steepening of the spectral index is observed with increasing source distance. This can be interpreted as absorption by the EBL. In general, knowledge of the intrinsic source spectrum is necessary to determine the density of the intervening EBL. Motivated by the observed spectral steepening with redshift, upper limits on the EBL are derived by assuming that the intrinsic spectra of the six blazars are $\propto E^{-1.8}$. Upper limits are then placed on the EBL flux at discrete energies without assuming a specific spectral shape for the EBL. This is an advantage over other methods since the EBL spectrum is uncertain.
\end{abstract}





\section{Introduction\label{sec:introduction}}
\begin{figure}
\epsscale{.7}
\plotone{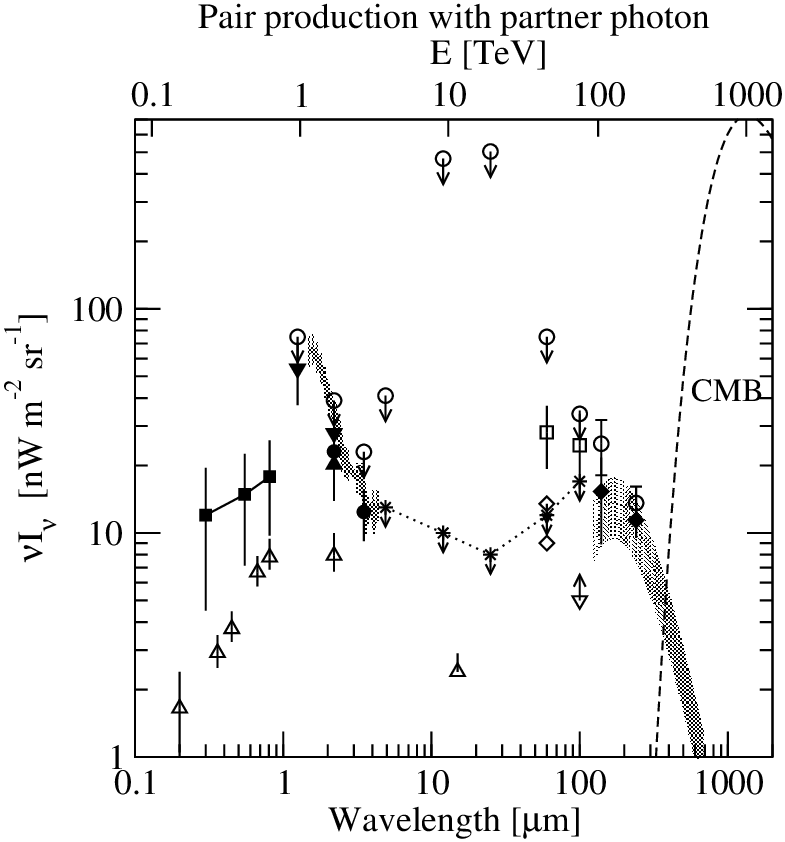}
\caption{Spectral energy distribution of the EBL. The top axis indicates the most likely energy of the partner photon to participate in pair-production with the EBL. The flux measurements and limits are taken from \citep{COBE_DIRBE_Hauser98} (\emph{open circles}),  \citep{IRB_60_100_with_DIRBE_Finkbeiner2000} (\emph{open squares}).
Upper limits from fluctuation analysis by \cite{CIBHauserDwek01} using the data from \cite{fluctuationCOBE_Kashlinsky96, FluctuationCOBE_Kashlinsky2000} (\emph{stars}). Tentative detection by \citep{60mu_Miville02} (\emph{open diamonds}).
\citep{COBE_FIRAS_Lagache99} (\emph{filled diamonds}),  \citep{COBE_DIRBE_cosmological_implications_Dwek98} (\emph{open triangle pointing down}),  \citep{FIRAS_spectrum_Fixsen98} (\emph{shaded region)} near the CMB. 
Galaxy counts corresponding to lower limits are shown with \emph{open triangles pointing up} and are at 0.2\mm\ from \citep{FarUV_EBL_Armand_94}, from 0.36 \mm\ to 2.2 \mm\ from  \citep{Hubble_deep_field_Pozzetti98}, and at 15 \mm\ from  \citep{CIB-ISOCAM_Elbaz02}.
In the infrared, the \emph{filled circles} are from  \citep{COBE_2.2_3.5_Wright2000}, \emph{filled triangle pointing up} from \citep{DIRBE-2MASS_Wright01}, and \emph{filled triangles pointing down} from \citep{CIB_at_1.25_2.2_Cambresky01}. In the optical, the \emph{filled squares} are from  \citep{HST_Campanas_1_Bernstein02}. The \emph{grey line} in the optical/infrared is from \citep{IRTS_Matsumoto2000}. 
The cosmic microwave background (CMB) is shown by a \emph{dashed line}.}
 \label{fig:ebl_ir}
\end{figure}

The measurement of the Extragalactic Background Light (EBL) is important for VHE gamma-ray astronomy as well as for astronomers modeling star formation and galaxy evolution. Second only in intensity to the Cosmic Microwave Background, the optical and infrared (IR) EBL contains the imprint of galaxy evolution since the Big Bang. This includes the light produced during formation and re-processing of stars. Current measurements of the EBL are summarized in Fig.~\ref{fig:ebl_ir}. The optical to near-IR emission, peaking in the 1 \mm\ region and extending to 20 \mm\, is due to direct star light, while molecular clouds and dust reprocess the optical light and emit in the mid to far IR region producing the second peak, $\lambda \approx$ 20-300 \mm. \citet{CIBHauserDwek01} comprehensively reviewed measurements and implications of the cosmic infrared background.

The optical to far-infrared EBL is difficult to measure because it is dwarfed by the much brighter foregrounds caused by night-sky glow, diffuse dust in the Galaxy, and the zodiacal light caused by interplanetary dust \citep{Diffuse_NSB_reference_Leinert98, CIBHauserDwek01}. For example, emission by the zodiacal dusk peaks in the 25 \mm\ region, orders of magnitude above the low EBL density in this waveband. In the case of ground- or rocket-based observations, instrumental emission also plays a significant role. This is complicated by the fact that the only characteristic upon which a detection of the EBL can be based is that it has to be distributed isotropically. These difficulties have precluded ground- and rocket-based measurements from detecting the EBL at all \citep{CIBHauserDwek01}.

The measured flux of VHE gamma rays is attenuated by pair production with optical/IR photons. On extragalactic distances, most of this absorption occurs by the EBL \citep{AbsorptionNikishov62}; interactions with stellar and Galactic material and optical/IR photons are negligible \citep{IR_determination_Dwek94}. Thus, if one somehow knows the initial gamma-ray flux, VHE astronomy is in a unique position to place limits on the density of the intervening optical/IR component of the EBL.

The organization of this paper is as follows: existing measurements and constraints on the EBL are presented in Sect.~\ref{sec:EBL_measurements}. Then, the brightest flare spectra from each of the six blazars are presented in Sect.~\ref{sec:comparison_of_spectra} together with the apparent spectral steepening with redshift. This is followed by a short review of pair-production absorption in Sect.~\ref{sec:theory} and is illustrated for the particular case of an assumed monoenergetic EBL in Sect.~\ref{sec:monoenergetic}. Upper limits on the EBL density are derived in Sect.~\ref{sec:upper_limits} and conclusions are given in Sect.~\ref{sec:conclusion}.
\section{Measurements and Constraints on the EBL\label{sec:EBL_measurements}}
Direct measurements are possible in the two windows of least foreground around 1 \mm\ and $>$100 \mm\  \citep{CIBHauserDwek01}. Recently, the Cosmic Background Explorer (COBE) satellite with its two instruments, the Diffuse Infrared Background Experiment (DIRBE) and the Far Infrared Spectrometer (FIRAS), has detected the EBL at 140 \mm\ and 240 \mm, see Fig.~\ref{fig:ebl_ir}. The possible detections at 60 \mm\ and 100 \mm\  \citep{IRB_60_100_with_DIRBE_Finkbeiner2000} are viewed as too high and are controversial, requiring revised galaxy evolution models with larger dust content \citep{60mu_EBL_AGN_Blain02}. The FIRAS measurement \citep{FIRAS_spectrum_Fixsen98} shows that the IR EBL can be characterized between 125 $\mu$m and 2000 mm by a modified blackbody spectrum. The isotropic optical and near-IR emission detected with the Japanese IRTS satellite is considerably higher than integrated light from galaxies and theoretical predictions \citep{IRTS_Matsumoto2000}.

Lower limits on the EBL density are placed by adding the flux per unit area received from all galaxies down to a given flux limit. As galaxies are only one source contributing to the EBL, these galaxy counts represent a lower limit on the total EBL \citep{galaxy_counts_Franceschini91, FarUV_EBL_Armand_94, Hubble_deep_field_Pozzetti98}. In the mid-IR region, where the foreground is particularly bright, \cite{CIB-ISOCAM_Elbaz02} were able to place a lower limit on the 15 \mm\ EBL density.

Upper limits can be placed on the EBL from direct measurements with minimal background subtraction \citep{COBE_DIRBE_Hauser98}. Also, upper limits can be derived from fluctuations in the measured light distribution, see \cite{fluctuationCOBE_Kashlinsky96, FluctuationCOBE_Kashlinsky2000}.


%
%
%
%
\subsection{Upper Limits from Observations of VHE Blazars}
BL Lacertae (BL Lac) objects are one type of active galactic nuclei that have a jet of high energy particles aligned with our line of sight. The nonthermal emission from the jet is highly variable and the mechanism for production of the VHE gamma-ray peak is under debate. In all models the intrinsic VHE spectrum of the source is smooth and concave downwards; no physical mechanism has been proposed that would produce an exponential rise with energy or emission-like line features.  The VHE spectrum measured on Earth is modified by interactions of gamma rays with the EBL. For example, a 1 TeV gamma ray reaches the threshold for electron-positron pair production if it collides with a 1 eV optical photon. If a pair is created, this gamma ray is lost and causes an attenuation of the measured spectrum at that energy. Thus, the EBL density could be inferred from the measured spectrum if the intrinsic spectrum of the blazar were known. In addition, because of the coarse energy resolution of this method, a measurement of the EBL spectrum is only possible if the shape of the EBL spectrum is known in advance.

Three methods have been used so far to derive EBL limits: 
\begin{enumerate}
\item Using multiwavelength observations, the intrinsic VHE blazar spectrum is predicted through interpolation or fitting of an emission model. The EBL spectral shape is usually fixed in advance with only the absolute normalization of the EBL flux left as a free parameter \citep{TeV_EBL_probe_Stecker92, upper_limits_Stecker93, Estimate_deJager94, IR_determination_Dwek94, Upper_limit_Biller95, Constraints_Guy2000}.
\item The deviation, or lack thereof, of the VHE spectrum from a predefined intrinsic spectrum, e.g. a power-law, is used to place upper limits on the EBL density with a given spectral shape \citep{Mrk501_EBL_Funk98, EBL-Mrk501_Stanev98, EBL_limits_Biller98, EBL_Vassiliev00}.
\item For a given EBL scenario the intrinsic source spectrum is reconstructed. EBL models are ruled out, or limits are placed on the density, if they give rise to an unphysical source spectrum: such as that it must not rise exponentially and that it should be consistent with the X-ray synchrotron peak \citep{time_averaged_Mrk501_Aharonian99, Constraints_Guy2000, Dust_Dwek00, EBL-Mrk501_Renault01, EBL_Krennrich03, EBL_Dwek_Krennrich04}.
\end{enumerate}

The method described in this paper differs from the above in that: (1) The intrinsic source spectrum of all detected VHE blazars is assumed to be identical to each other. A conservative upper limit to the intrinsic spectrum is interpolated from the observed correlation between the spectral index of the source during a flare state and its redshift. (2) Upper limits to the EBL intensity are calculated only at single energy points where the VHE spectrum has been measured, avoiding the need to specify the EBL spectral shape.

\section{Comparison of the Brightest Flare Spectra\label{sec:comparison_of_spectra}}
VHE emission has been detected and confirmed from six BL Lac objects: Mrk~421~\citep{Mrk421_Punch92,Mrk421_Petry96}, Mrk~501 \citep{Mrk501_Quinn96, Mrk501_Bradbury97}, 1ES~2344+514 \citep{2344Catanese98, 2344Hegra03}, 1ES~1959+650 \citep{1es1959Nishiyama99, 1es1959Holder03, 1es1959Aharonian03}, PKS~2155-304 \citep{pks2155_chadwick99, HESS_astroph_2155}, and H~1426+428 \citep{H1426_horan02, HEGRA_H1426_Aharonian02}.

For all six blazars, hysteresis and the rarity of simultaneous X-ray/VHE spectra make it difficult to establish a temporal correlation between X-rays and VHE gamma rays. The sources are similar in that the X-ray synchrotron peak ranges from $10^{17}$ to almost $10^{20}$ Hz with peak luminosities between $10^{43}$ and $10^{44}$ erg s$^{-1}$ sr$^{-1}$ \citep{multi_wave_1es1959_Krawczynski04}. The black hole mass estimates are also relatively similar: $10^{8}$-$10^{9}$ M$_\odot$. It is possible that the reason for this similarity lies in the fact that only the brightest VHE blazars are seen by the imaging atmospheric Cherenkov telescopes at this time and that there is actually a continuous population of blazar spectra \citep{stretching_Costamante01}. 

The brightest flare spectra measured from each of the six blazars are shown in Fig.~\ref{fig:spectra_AGN}; they are increasingly steep with source distance. The spectra for Mrk~421 \citep{Mrk421_Krennrich01} and Mrk~501 \citep{Mrk501_Samuelson98} are best described by a power law with exponential cut-off \citep{Mrk421_Krennrich01}. The cut-off, at energy of about 4 TeV, can be attributed to gamma-ray absorption by the extragalactic medium.  For 1ES~1959+650, the gamma-ray 'orphan' flare is used \citep{thesis_Schroedter04}, a TeV flare during which the X-ray spectrum was not particularly bright \citep{multi_wave_1es1959_Krawczynski04}. For 1ES~2344+514, the brightest flare recorded thus far was recorded in 1995 by the Whipple 10 m telescope; the spectrum has been derived in \cite{thesis_Schroedter04, 2344_Schroedter05}. The spectrum of PKS~2155-304 has recently been measured by the HESS collaboration with high precision \citep{HESS_astroph_2155}. Though the light curve shows variability, the spectrum can be well fitted by a single power law over the entire energy range while the power law index did not change with flux level. VHE emission from the furthest object, H~1426, is generally very low \citep{H1426_Falcone04}, so that no flare spectra have been measured. However, as VHE emission during flaring episodes is much higher than during the quiescent state, most of the photons in the spectrum by \citep{spectrum_H1426_Petry02, HEGRA_H1426_Aharonian03} may be associated with flares.

\begin{figure}
\epsscale{1}
\plotone{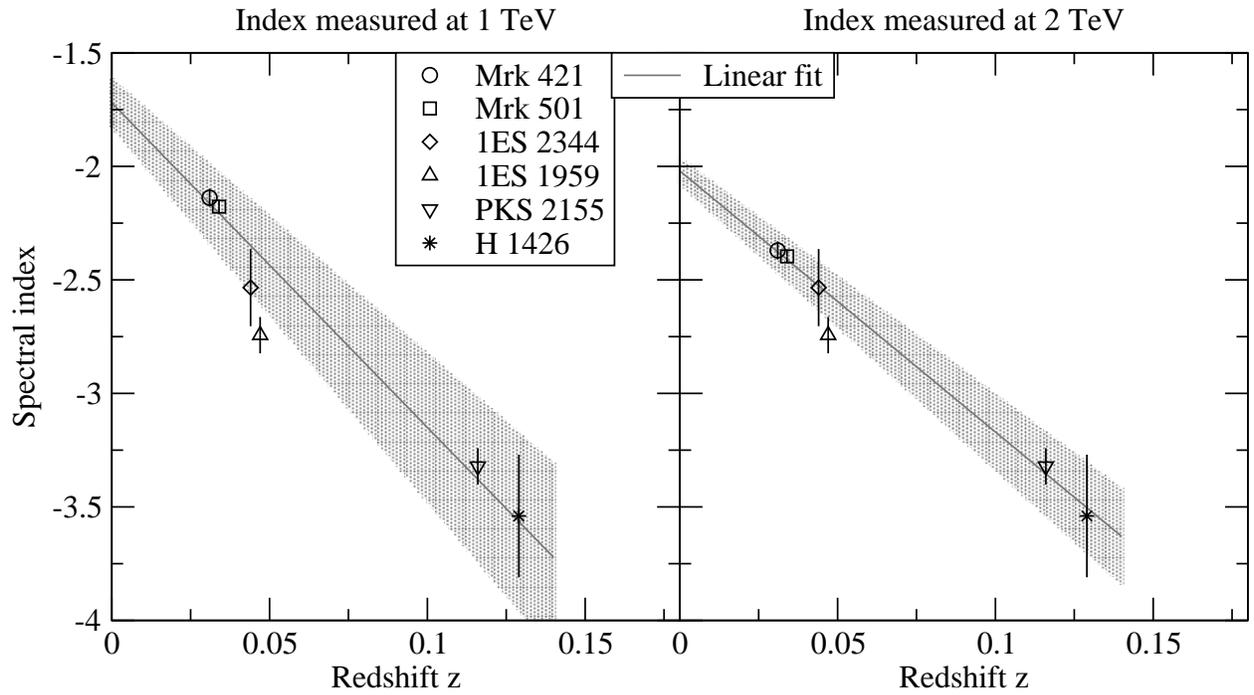}
  \caption{Spectral index of power law fits to the hardest measured flare spectra versus distance of the VHE blazars. The index was measured at 1 TeV (\emph{left}) and at 2 TeV (\emph{right}) for the curved spectra of Mrk~421 and Mrk~501. A linear fit to the data is shown in each case.}
 \label{fig:z_index}
\end{figure}

The power-law spectral index fitted to each AGN flare spectrum is shown in Fig.~\ref{fig:z_index}. Because a significant cutoff is present in the spectra of Mrk~421 and Mrk~501, the spectral index is measured at two energies, 1 and 2 TeV, to estimate the systematic error arising from this choice. In both cases, the increasingly steep spectral index with redshift was fitted with a straight line, resulting in a spectral index at $z=0$ of -1.7$\pm$0.1 with $\chi^2$/(d.o.f.) = 22.3/4 when the spectral index is measured at 1 TeV and index of 2.0$\pm$0.1 ($\chi^2$/(d.o.f.)=5.7/4)  when the index is measured at 2 TeV. The flare spectra, after taking into account the redshift dependence, appear similar and hence we make the zeroth order assumption that the same process is responsible for the VHE emission in all these AGN.

In the interpretation that the cut-off in the energy spectrum of Mrk~421 and Mrk~501 is due to absorption by the EBL, note that for a power law fit with exponential cut-off, the power law index ranges from -1.9 to -2.1, with a cut-off energy around 4 TeV consistent for both blazars and independent of flux state \citep{spect_var_Mrk421_Krennrich02, spect_var_Mrk421_Aharonian02}.

%
%
%
%
%
\subsection{Spectral Steepening Predicted by an EBL Model\label{sec:EBL_models}}
The literature contains various EBL models with various merits; a large fraction has been combined into different EBL scenarios by \citet{EBL_Dwek_Krennrich04}. All of the scenarios that do not imply unreasonable intrinsic VHE spectra and most of the recent EBL models, produce similar gamma-ray opacities. Here, one particular EBL model is used as a reference, put forward by \citet{model_Primack} and referred to as the Primack model in the following. This model is compared in Fig.~\ref{fig:ebl_models} with EBL measurements and with the allowed EBL scenarios of \citep{EBL_Dwek_Krennrich04}. Although \citeauthor{model_Primack} has proposed a number of models and the one chosen here is known to have slight problems above 60 \mm, it will be used for illustrative purposes only, with the exception of giving an estimate of the optical depth already present for low energy gamma rays in Sect.~\ref{sec:upper_limits}.

%

\begin{figure}
\epsscale{.6}
\plotone{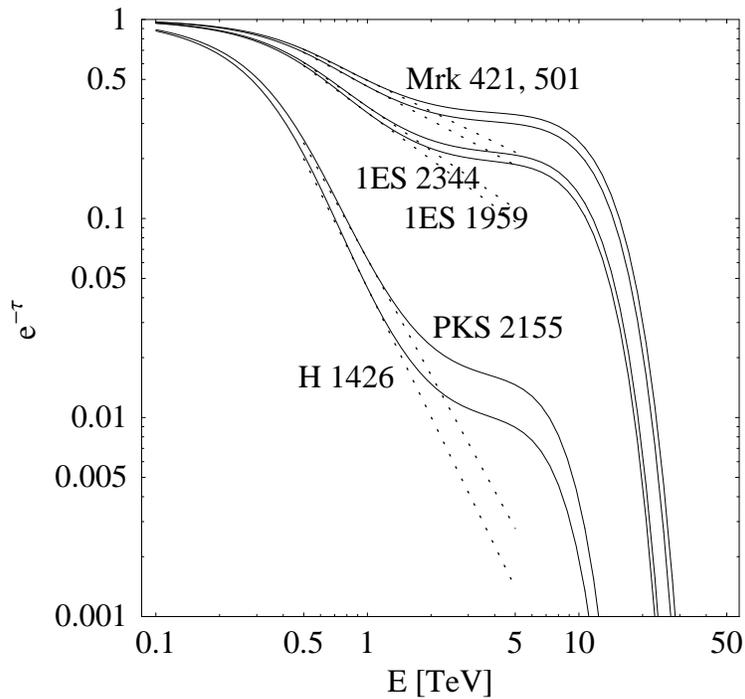}
\caption{Illustration of power law spectra measured at 1 TeV from the attenuation predicted by the Primack model. The attenuation, $e^{-\tau}$ (\emph{solid lines}), is shown for the blazars (\emph{top to bottom, solid lines}) Mrk~421, Mrk~501, 1ES~2344, 1ES~1959, PKS~2155, and H~1426. The \emph{dashed lines} show the power-law spectra at 1 TeV. These become steeper with increasing source distance. No attenuation occurs at $z=0$, the horizontal line at the top of the figure, $e^{-\tau}=1$, corresponds to the intrinsic source spectrum and could have any reasonable shape. }
 \label{fig:primack_power_fit}
\end{figure}

The attenuation of the intrinsic spectrum in the Primack EBL model is shown in Fig.~\ref{fig:primack_power_fit} at the distances of the six blazars. The figure shows the spectral slope, measured at 1 TeV, becoming increasingly steep with redshift. The steepening of blazar spectra with redshift was predicted by \citet{Steepening_Stecker99}. In the case of PKS~2155, the absorbed spectrum could be fitted with an $E^{-3}$ spectrum over the energy region from 0.3 TeV to 3 TeV, assuming an initial $E^{-2}$ spectrum.
%
%

\begin{figure}
\epsscale{1}
\plotone{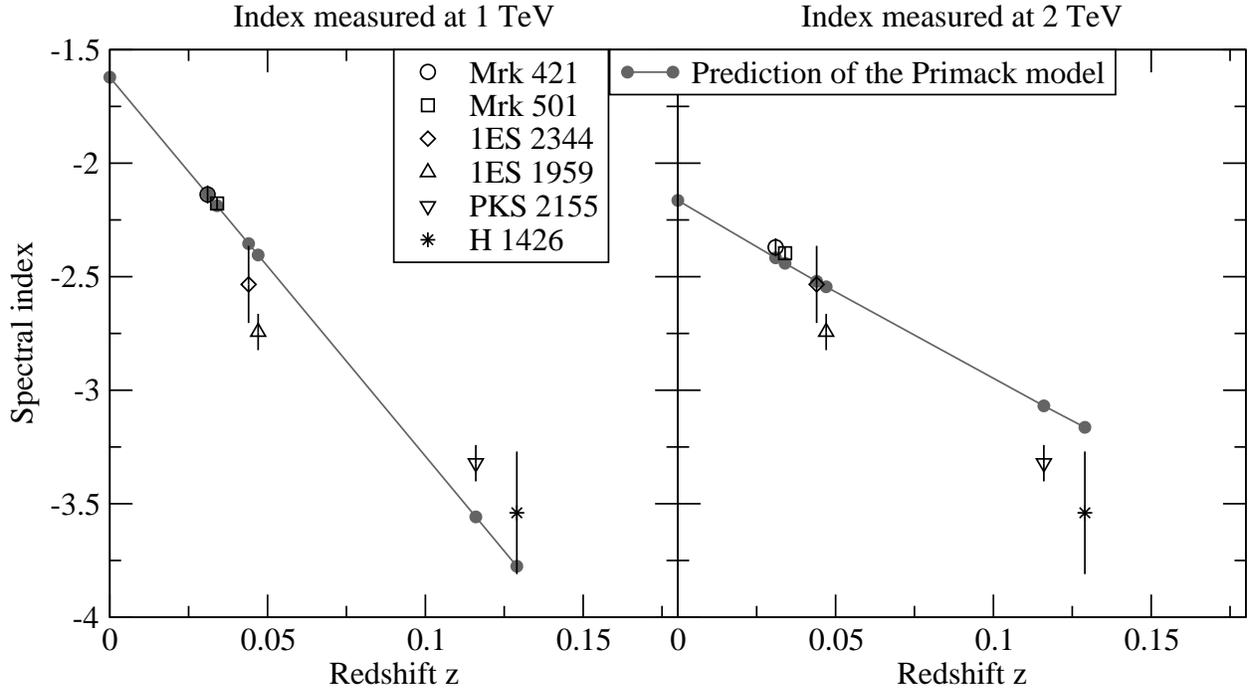}
  \caption{Spectral index of power law fits to the hardest measured flare spectra versus distance of the VHE blazars. The index was measured at 1 TeV (\emph{left}) and at 2 TeV (\emph{right}) for the curved spectra of Mrk~421 and Mrk~501. The prediction of the Primack model for the steepening of the spectra is shown by \emph{line with points}. The predicted index was evaluated at each source distance and produces an almost linear relationship. For the Primack model, the spectral index at zero redshift is a free parameter and is derived from a best-fit to the data. The reduced $\chi^2$ for the fit in the \emph{left} figure is 5.7 and 4.3 for the \emph{right} figure. }
 \label{fig:primack_z_index}
\end{figure}

A linear fit to the spectral index steepened by the Primack model, measured at 1 and 2 TeV, vs. redshift, is shown in Fig.~\ref{fig:primack_z_index}. To determine the intercept of the linear fit, i.e. the intrinsic source spectral index, we assume that all blazars have the same intrinsic spectrum and adjust the level until the $\chi^2$ difference between measured and predicted indexes is minimized. The spectral index at zero redshift predicted in this way is -1.62 and -2.2, when the power law fits are performed at 1 TeV and 2 TeV, respectively. Considering the large degree of uncertainty in choosing an energy at which to measure the spectral index, Fig.~\ref{fig:primack_power_fit} implies that the intrinsic spectral index is on the order of -2 to -1.8. 

It should be noted that the detailed shape of the spectral steepening with increasing redshift depends strongly on the shape of the EBL. For example, the seven possible EBL scenarios given in \cite{EBL_Dwek_Krennrich04} allow a range of slopes for spectral steepening with redshift, with the Primack model roughly representing the average of the allowed EBL scenarios.

Fig.~\ref{fig:primack_power_fit} shows that in the Primack EBL model, there is a leveling-off at high energies, not a simple power law with exponential cut-off as supposed above. The cut-off energy in this case would correspond to the initial downturn at $\approx$ 0.6 TeV. At higher energies, the flux is so far diminished that measurements with VHE telescopes do not yield a statistically significant result. Note that if the optical depth is independent of energy, as is the case for $\nu F_{\nu}(\lambda) \propto \lambda^{-1}$, no change occurs at all in the spectrum with increasing redshift. In the Primack model this is almost the case between 2 TeV and 5 TeV, corresponding to the ~2 - 5 \mm\ EBL.
\section{Attenuation due to Pair Production\label{sec:theory}}
Consider a single VHE gamma ray of energy, $E$, coming from a distant extragalactic source. If the gamma ray makes a collision with another particle, it will not propagate in a straight line and is lost to the observer who is several Mpc away. Pair production is the most likely type of inelastic collision in extragalactic space as the photon density, though varying with wavelength, is much higher than the matter density. For pair production to occur, the total energy available must be greater than $2 m_e$. For VHE gamma rays with energy between 100 GeV and 20 TeV, the low energy photon must be in the range from 10 eV to 0.05 eV. The absorption of VHE gamma rays is described by a very simple rate equation $\gamma_{TeV} + \gamma_{ir} \rightarrow e^+ + e^-$. The number of VHE gamma rays, $N$, with energy $E$ changes per unit time, $\ud t$, as
\begin{equation}
\label{eq:population}
\frac{\ud N(E)}{\ud t} = - N(E) \lambda_{\gamma\gamma\rightarrow e^+ e^-}(E),
\end{equation}
where $\lambda$ is the reaction rate defined below. Eq.~\ref{eq:population} is solved by expressing $\ud t$ in terms of the distance $\ud l$ traveled by the photon: $\ud l / \ud t=c$. For small redshift, $z<<1$, the relation $\ud l / \ud z=c/H_0$ holds so that the number of gamma rays at redshift $z$ is given by
\begin{equation}
N(z,E)=N_0(E)\ e^{- H_0^{-1} z\ \lambda(E)}.
\label{eq:absorbed_N}
\end{equation}
This defines the optical depth 
\begin{equation}
\tau(z,E) = H_0^{-1} z\ \lambda(E),
\label{eq:tau}
\end{equation}
with $H_0=71\pm4$ km s$^{-1}$ Mpc$^{-1}$ \citep{WMAPBennett03}. This formalism can been extended to cosmological distances \citep{absorption_lowE_Salamon98, modified_AGN_spectra_Biller95}. The optical depth of the furthest detected VHE blazar H~1426 at $z=0.129$, calculated according to Eq.~\ref{eq:tau} and with the cosmologically correct version differs by less than 10\%. As this difference is less than the systematic uncertainty of VHE blazar spectra, Eq.~\ref{eq:tau} will be used. The measured flux, $F_m$, is then given in terms of the emitted flux, $F_e$, by $e^{-\tau(E_m)} \ F_e( E_m )$, where $E_m$ is the measured energy. This ignores the inverse-square law distance dependence; it is of no concern as it effects only the absolute flux level and does not have an energy dependence.

The momentum distribution averaged pair production rate, $\lambda$, in units of [$T^{-1}$], is
\begin{eqnarray}
\lambda(E) \equiv \langle \sigma v_{rel} \rangle & = & \int \ud^3 p\  f(p)\ \sigma(\sqrt{s})\ v_{rel} \\
& = & \int \ud\Omega\ \int\ p^2\ \ud p\  f(p)\ \sigma(\sqrt{s})\ v_{rel} \\
& = & 2\pi c \int_{-1}^{+1} \ud x\ (1-x) \int_{\frac{2 m^2}{E(1-x)}}^\infty p^2\ \ud p\ f(p)\ \sigma(\sqrt{s}),
\label{eq:lambda}
\end{eqnarray}
where $\ud\Omega \equiv \ud(cos \theta)\ud \phi$, $cos \theta \equiv x$ is the angle between the incoming particles, $\phi$ is the azimuthal angle, and $\sqrt{s}$ is the total energy. The relative speed between the interacting particles is $v_{rel}=c (1-x)$. The comoving photon momentum distribution of the EBL is given by $f(p)$ [$E^{-3} L^{-3} sr^{-1}$]. Eq.~\ref{eq:lambda} can also be written in the more familiar way of particle density per energy per volume by substituting $n(\epsilon)\equiv 4\pi p^2 f(p)$ and with the replacement $p \rightarrow \epsilon$. 
The spin-averaged pair production cross section, $\sigma$, is given by \citep{PairBreit34}. From Eq.~\ref{eq:lambda} it can be seen that the optical depth is independent of the gamma-ray energy  if the EBL photon density, $n(\epsilon)$, is independent of $\epsilon$: i.e. $\lambda(E) \propto n$ only. This corresponds to an energy density $\ud n/\ud \epsilon  \propto \epsilon^{-1}$.   On a $\nu F_{\nu}$ plot, such an EBL spectrum would fall as $\lambda^{-1}$ and is quite possible in the optical / near-IR portion.
\subsection{Upper Limits from the Assumption of a Monoenergetic EBL\label{sec:monoenergetic}}
\begin{figure}
\epsscale{.5}
\plotone{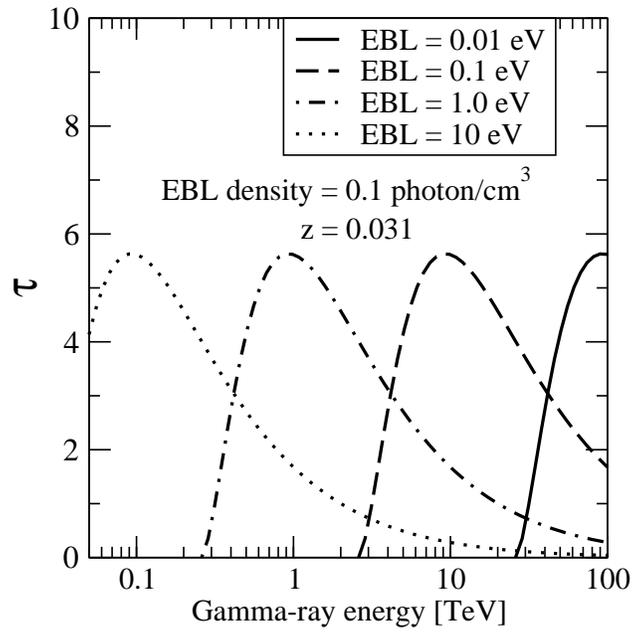}
  \caption{Absorption probability of a VHE gamma ray by a monoenergetic EBL of energy $\epsilon$ = 0.01 eV (123 $\mu$m, solid), 0.1 eV (12.3 $\mu$m, dashed), 1.0 eV (1.23 $\mu$m, dash-dot-dot), and 10 eV (123 nm, dotted). The EBL photon density is 1/cm$^3$.}
 \label{fig:line_depth}
\end{figure}

The sensitivity of the VHE gamma-ray spectrum to absorption by the EBL is best illustrated by the toy model of a monoenergetic EBL, i.e. isotropically distributed photons with only a single energy. The absorption probability per unit length is given by
\begin{equation}
\lambda(E, \epsilon) = n(\epsilon) \frac{c}{2} \int_{-1}^{+1} \ud x\ (1-x)\ \sigma(v_e),
\label{eq:monoenergetic_rate}
\end{equation}
where the constant $n(\epsilon)$ is the co-moving photon density [$L^{-3}$]. Note that the only dependence on the EBL energy $\epsilon$ is through $v_e$. The absorption probability per unit length is plotted in Fig.~\ref{fig:line_depth} for various EBL energies ranging from IR to UV: the density is taken uniformly as 1 cm$^{-3}$. Though gamma-rays are absorbed most efficiently when $E \epsilon=0.93\times 10^{12}$~eV$^2$, there is significant absorption of gamma rays with energies from half to four times as much.

In general, knowledge of the optical depth is not sufficient to unambiguously determine both the shape and magnitude of the EBL density. If the EBL flux is known over a finite wavelength region, an infinite number of shapes are possible because of the limited energy resolution of this method. The energy resolution is limited by the width of the pair production cross-section and the isotropic EBL photon distribution, see Eq.~\ref{eq:lambda}. If one assumes a shape for the EBL spectrum with only the overall normalization left as a free parameter, it is possible to determine the best fit EBL flux through a $\chi^2$ minimization between the measured and modeled optical depths. However, this still relies on knowing the intrinsic blazar spectrum, which is not easily accomplished. This method has been widely used to derive upper limits for power-law EBL spectra.

To avoid specifying the EBL spectral shape altogether, upper limits on the true EBL density are derived here by assuming the EBL to be monochromatic. Physically, we know that the EBL spectrum is extended. If we make that assumption that the EBL is monochromatic, the contribution of other wavelengths to the absorption of gamma rays is ignored. Hence, the EBL density at that single wavelength will have to be larger to reproduce the optical depth, than if other wavelengths were allowed to contribute to the absorption.

\begin{figure}
\epsscale{1}
\plotone{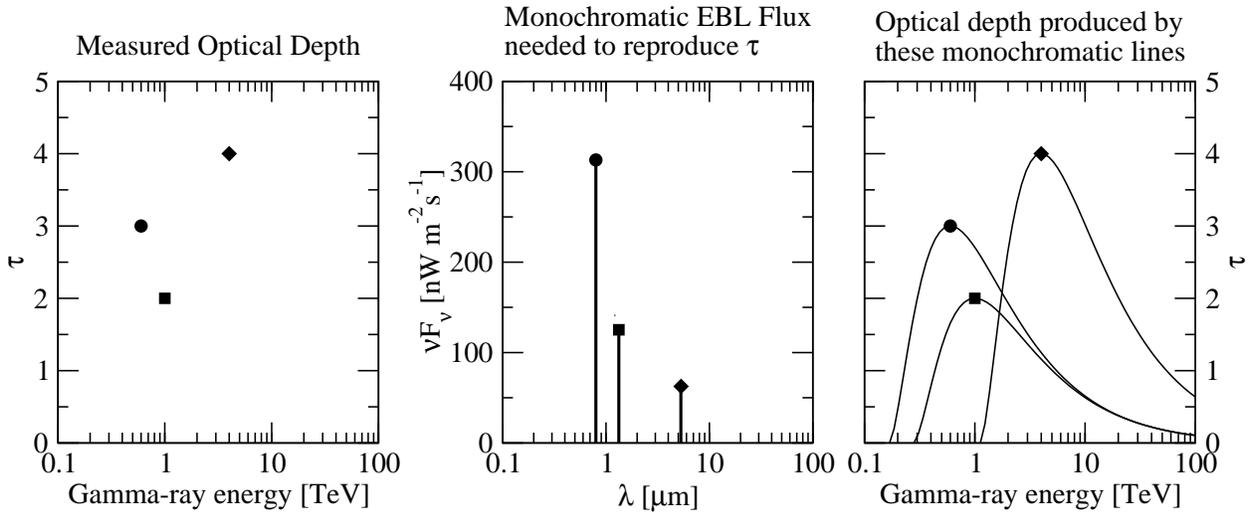}
  \caption{Illustration of the method to derive an upper limit on the EBL density from a measurement of the optical depth. From the measured optical depth (\emph{left}), the monochromatic EBL intensity is calculated (\emph{middle}). The (\emph{right}) graph shows the optical depth produced by the monochromatic EBL line. Note that the optical depth derived from each monochromatic line overestimates the measured optical depth everywhere except at the measured gamma-ray energy. The total absorption is the sum of the three lines.}
 \label{fig:tau_limit_tau}
\end{figure}

Suppose we start with a known measurement of $\tau(E)$ from a source at $z=0.031$, illustrated in Fig.~\ref{fig:tau_limit_tau} (\emph{left}) for three arbitrarily chosen values of $\tau(E)$. Then, we want to know: What EBL density does it take to reproduce this amount of absorption? The answer to this question depends on what photon wavelength is performing the absorption. The least EBL density is required if it absorbs the gamma-ray most efficiently, that is, when $E \epsilon=0.93\times 10^{12}$~eV$^2$. The monochromatic EBL density corresponding to this case can be expressed in terms of the measured optical depth, $\tau_m$, using Eq.~\ref{eq:tau}
\begin{eqnarray}
\frac{\tau_m}{z H_0^{-1}} & = & n(\epsilon) \frac{c}{2} \int_{-1}^{+1} \ud x\ (1-x)\ \sigma(v_e)\\
\rightarrow  n(\epsilon) & = & \frac{   \frac{\tau_m H_0}{z}   }{\frac{c}{2} \int_{-1}^{+1} \ud x\ (1-x)\ \sigma(v_e)},
\end{eqnarray}
so that the upper limit (UL) on the true density is given by
\begin{equation}
n_{\mathrm{UL}}(   \frac{0.93\ \mathrm{eV\ TeV}}{E}   ) = \frac{  \frac{\tau_m(E)\ H_0}{z\ c}  }{1.4\times 10^{-29} \mathrm{m}^2}.
\label{eq:upper_limit}
\end{equation}
Fig.~\ref{fig:tau_limit_tau} (\emph{middle}) shows the EBL flux corresponding to this case and Fig.~\ref{fig:tau_limit_tau} (\emph{right}) shows the corresponding optical depth produced by this single monochromatic EBL line. 

Consider then the case of many adjacent optical depth measurements, shown in Fig.~\ref{fig:tau_limit_tau} (\emph{left}) at only three energies. As before, the monochromatic EBL flux is calculated in turn from each optical depth (\emph{middle}). The total absorption, the sum of the three optical depths, is larger than was measured at the three energies (\emph{right}). Hence, the monochromatic EBL flux is an overestimate of the true value required to achieve the measured attenuation.
\section{EBL Upper Limits from Blazar Spectra\label{sec:upper_limits}}
Determination of the amount of absorption present in one measured spectrum requires a priori knowledge of the intrinsic source spectrum. Though the mechanism for VHE gamma-ray production has been modeled, which model to choose is still unknown. In addition, other important details are left open as well: What is the electron or proton spectrum? What is the strength of the magnetic field? What is the opacity in the vicinity of the source for gamma-rays to escape?

To place upper limits on the EBL density, the following simple assumption will be made: The intrinsic spectrum of all blazars are identical to a power law with index -1.8; this has been motivated in Sect.~\ref{sec:comparison_of_spectra}. If the VHE spectrum has not been measured down to 0.2 TeV, the flux at that energy is predicted from a power law fit extrapolated down from higher energies. The small amount of EBL absorption present already at 0.2 TeV is calculated from the Primack model. 

The assumed spectral index of -1.8 is very hard and cannot extend indefinitely. By using a power law fit from higher energies, where the measured spectrum is steeper due to absorption, to extrapolate down in energy to 0.2 TeV will likely overestimate the expected flux from the source. The assumed intrinsic spectral index is approximately implied by the physics of TeV blazars. The intrinsic spectra of TeV blazars are expected to peak in the multi-TeV region \citep{Predicted_Stecker96}; hence the intrinsic spectral index should be somewhat harder than -2. Detailed modeling show this to be the case for Mrk~421 and Mrk~501 \citep{modeling_Krawczynski02, Modeling_Konopelko03}. \citet{EBL_Dwek_Krennrich04} show that after accounting for absorption by the EBL, the intrinsic spectra, in a $\nu F_{\nu}$ representation, peak between 0.5-1.2 TeV for Mrk~421, between 0.8-2.5 TeV for Mrk~501, and between ~1-5 TeV for H~1426. 

\begin{figure}
\plotone{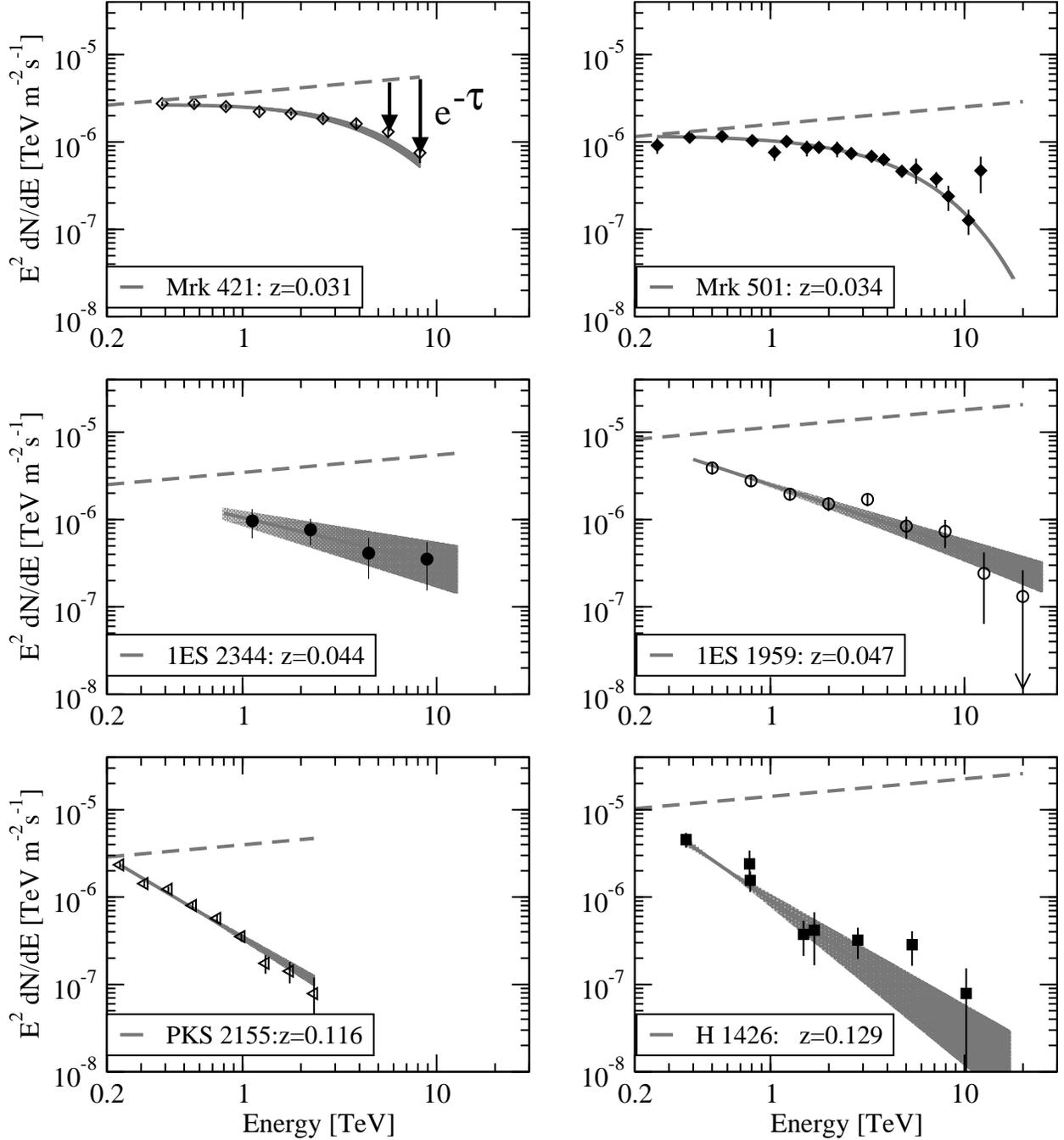}
  \caption{The very high energy flare spectra of six AGN. The flux of each source is adjusted with an inverse square law to the same distance as Mrk~421. Shown by the \emph{dashed line} in each case is the assumed source spectrum with differential index -1.8; they are normalized to the measured, or predicted, flux at 0.2 TeV. The \emph{top left} graph indicates with an arrow the effect of the optical depth in attenuating the flux at two energies. }
 \label{fig:spectra_AGN_E2}
\end{figure}
\begin{figure}
\epsscale{.9}
\plotone{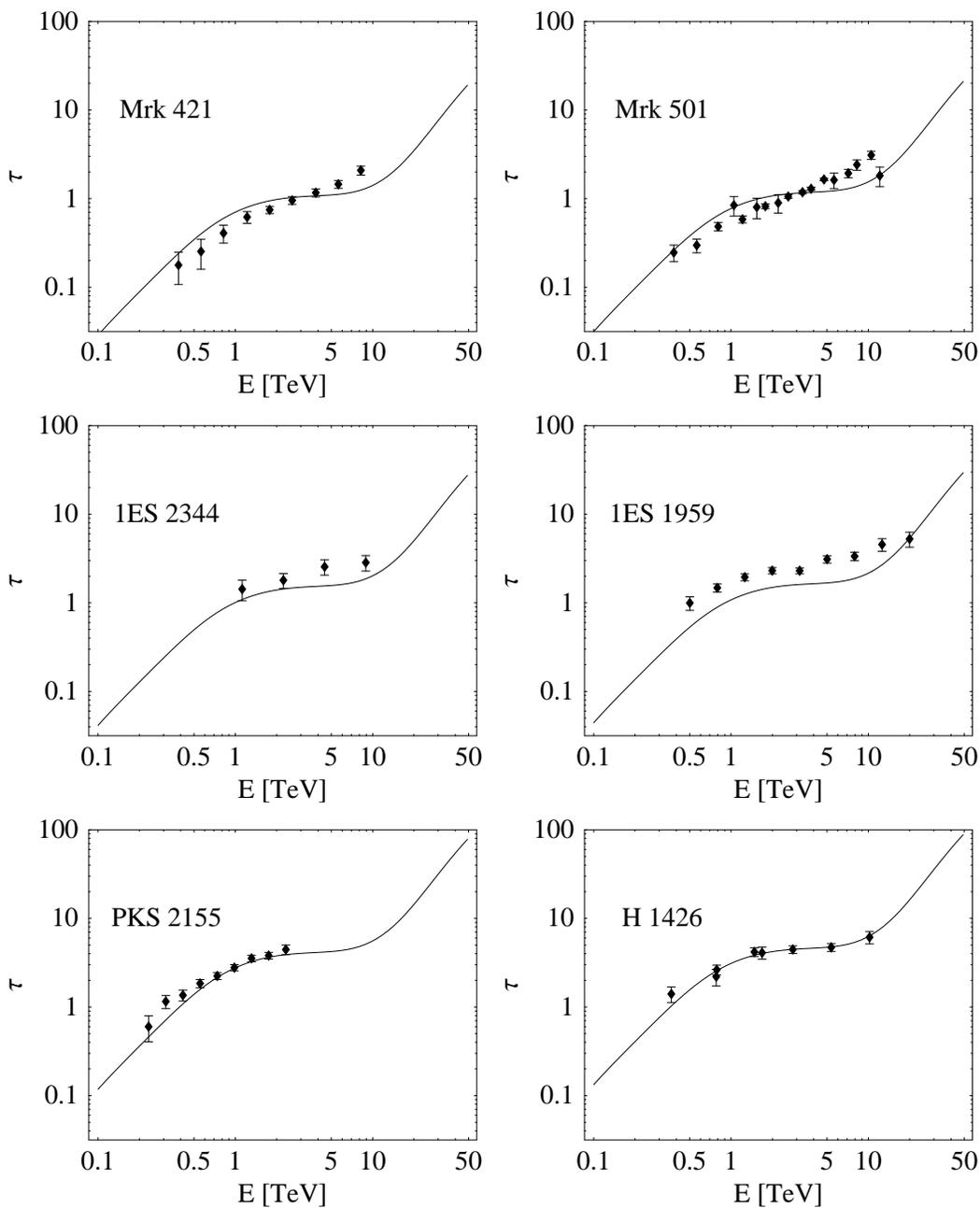}
  \caption{The optical depth (\emph{data points with error bars}) derived according to Eq.~\ref{eq:measured_depth} from the differences in Fig.~\ref{fig:spectra_AGN_E2} between measured and assumed flux. For comparison, the optical depth derived from the Primack model is shown by a \emph{solid line}. }
 \label{fig:tau_primack_data}
\end{figure}
\begin{figure}
\epsscale{.6}
\plotone{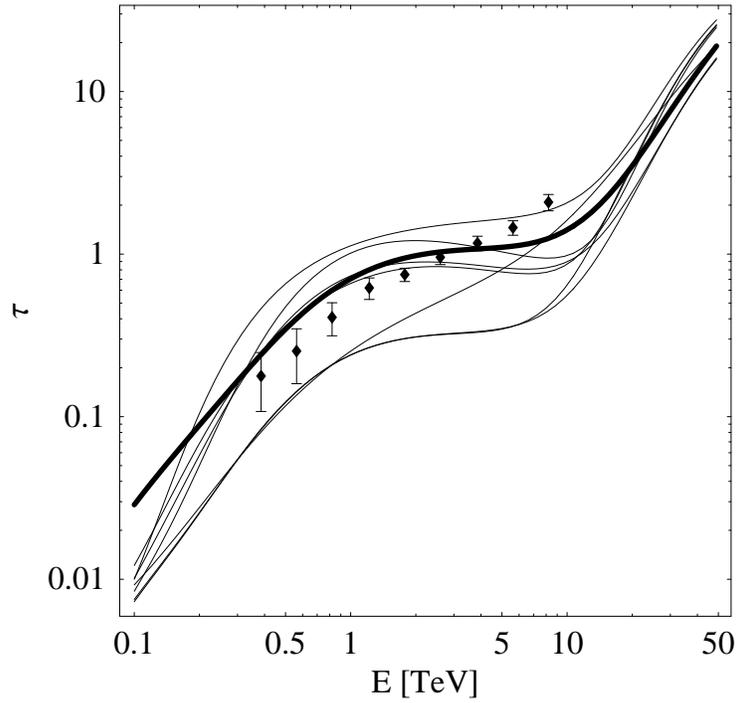}
  \caption{The optical depth (\emph{data points with error bars}) derived according to Eq.~\ref{eq:measured_depth} for Mrk~421. Superimposed are the optical depth derived from the Primack model (\emph{thick line}) and optical depths derived from the allowed EBL scenarios given in \citet{EBL_Dwek_Krennrich04} \emph{thin lines}. }
 \label{fig:tau_primack_dwek_mrk421}
\end{figure}

The expected flux in the absence of EBL absorption from each blazar is shown in Fig.~\ref{fig:spectra_AGN_E2} by a dashed line. The measured flux levels have been adjusted to the distance of Mrk~421 using an inverse square law to show that the power output of the sources could be similar if the sources have identical geometries. The optical depth is the ratio of expected flux to measured flux with the addition of the optical depth already present at 0.2 TeV
\begin{equation}
\tau = \ln{(F_e/F_m)} + \tau(z, E=0.2\ \mathrm{TeV})
\label{eq:measured_depth}
\end{equation}
and this does not dependent on the overall normalization of the flux level. The optical depth calculated in this way for the six blazars is shown in Fig.~\ref{fig:tau_primack_data}. Also shown in the figure is the optical depth calculated in the Primack model for comparison with the measured values. A comparison of the measured optical depth for Mrk~421 with the allowed EBL scenarios of \citep{EBL_Dwek_Krennrich04} is given in Fig.~\ref{fig:tau_primack_dwek_mrk421}. Here, the Primack model lies roughly in the middle of the allowed EBL scenarios. The measured optical depths of Mrk~421 and Mrk~501 are steeper in the 1-5 TeV region than some of the EBL scenario and would favor a larger EBL density in the mid-IR region, as in the IR spectrum derived by \citet{Malkan2001_EmpiricalModel}. The measured optical depths of the other sources are flatter, showing that this sample of six blazars is not completely homogeneous.

\begin{figure}
\epsscale{.9}
\plotone{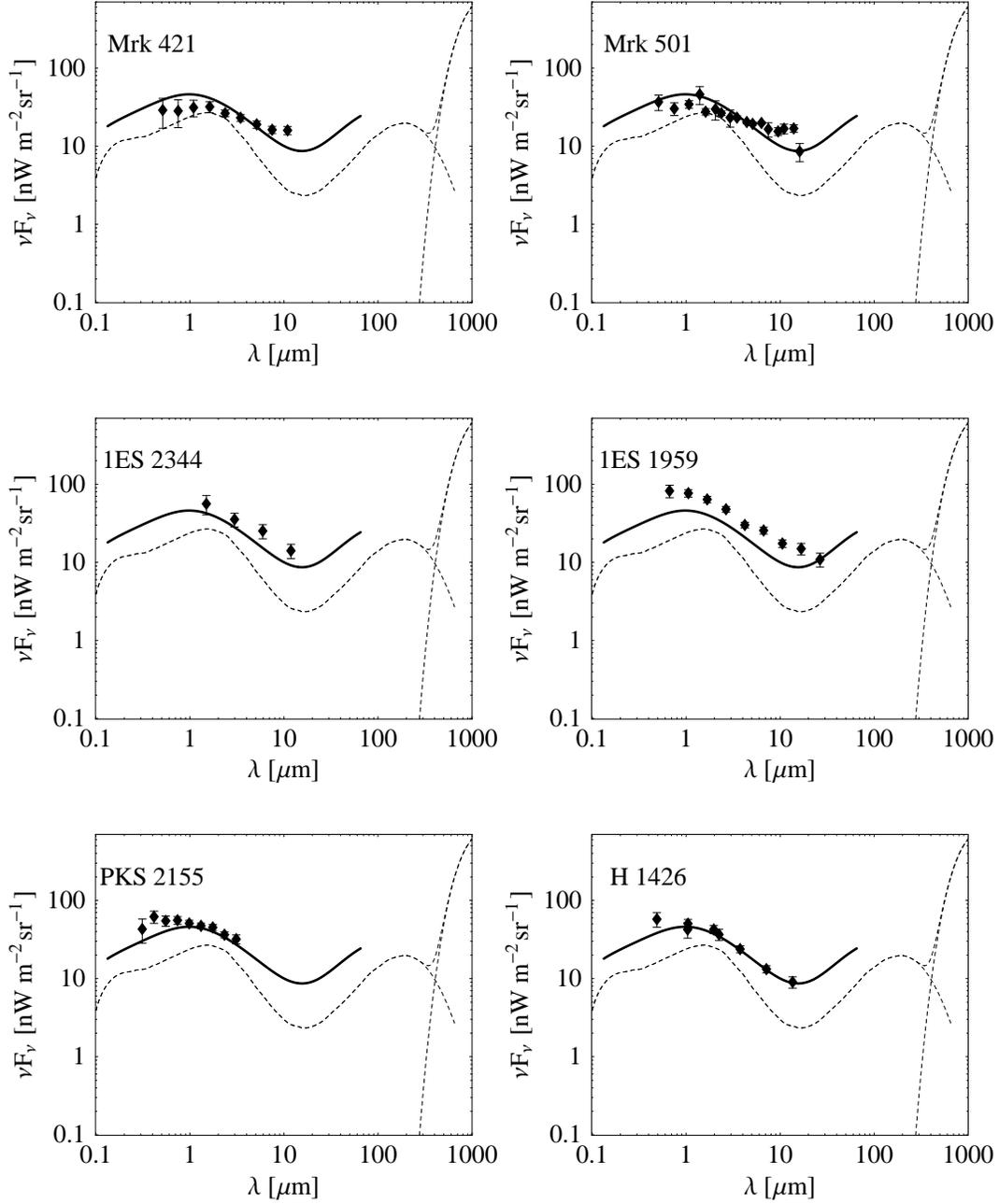}
  \caption{The monoenergetic EBL flux derived from the spectra of AGN (\emph{points with error bars}). Also shown for comparison is the monochromatic flux (\emph{thick line}) derived from the Primack EBL model (\emph{thin dashed line}).}
 \label{fig:EBL_upper_limits}
\end{figure}

The monochromatic EBL flux derived from each blazar spectral energy point under the above assumptions is shown in Fig.~\ref{fig:EBL_upper_limits}. For comparison, the value of the monochromatic flux as calculated from the Primack model is shown in the figure as well. For the most part, the monochromatic flux values derived here agrees with the Primack model; this is also true for all of the other reasonable EBL scenarios.

Differences in the derived monochromatic flux are likely caused by the intrinsic flare spectra of the sources being different from each other and not being pure power laws. Because of the large distances to the blazars, it is highly unlikely that non-uniformities in the EBL give rise to the observed differences unless an IR galaxy is in the line of sight. The limits derived from Mrk~421 and Mrk~501 are the lowest. To bring them higher and in line with the other sources, their intrinsic spectra would need to be slightly harder than -1.8 in the low energy region. If the spectra were harder then curvature with a peak in the ~1-2 TeV range would make the flux limits more consistent with those from the other sources. The systematic uncertainty in absolute energy calibration is 10\%, directly corresponding to a 10\% uncertainty in the EBL wavelength in Fig.~\ref{fig:EBL_upper_limits}.

\begin{figure}
\epsscale{.7}
\plotone{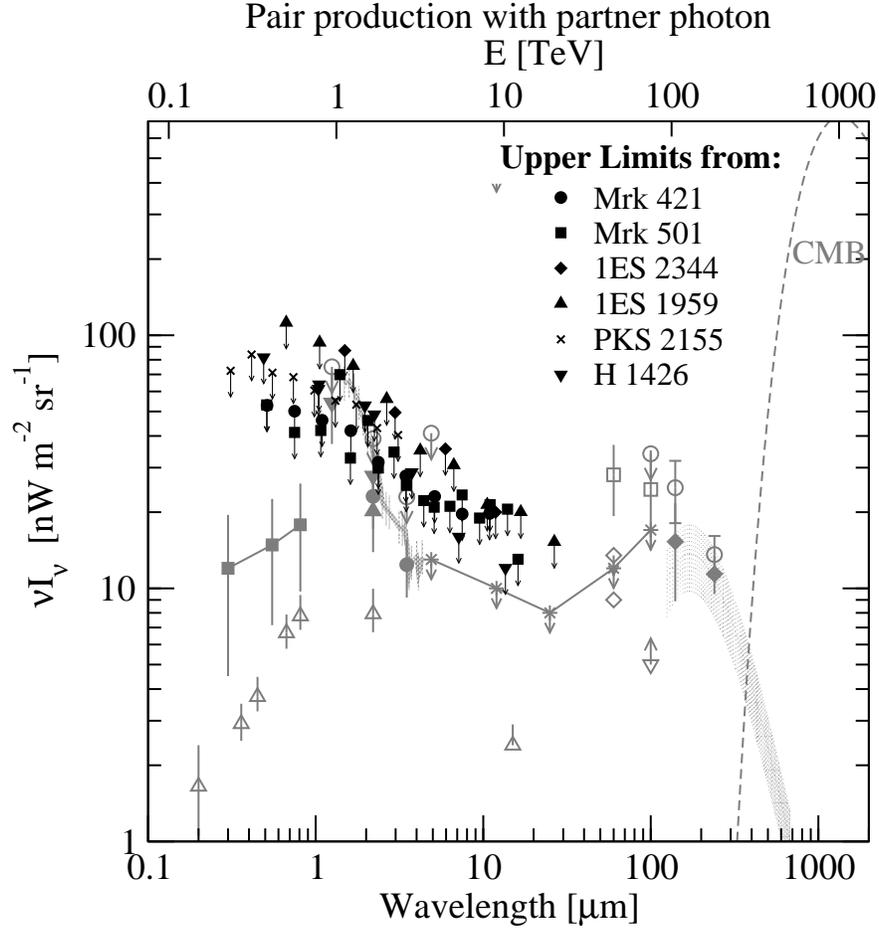}
  \caption{98\% confidence level upper limits of the EBL flux (\emph{points with upper limit error}). Measurements and limits from other publications are shown in \emph{gray}, see also Fig.~\ref{fig:ebl_ir}. The top axis indicates the most likely energy of the partner photon to participate in pair-production with the EBL. }
 \label{fig:ebl_limits}
\end{figure}

Fig.~\ref{fig:ebl_limits} shows the derived 98\% confidence level upper limits together with optical EBL measurements and limits shown previously in Fig.~\ref{fig:ebl_ir}. The range of upper limits from the six sources is indicative of the systematic error of this method. The systematic error stemming from the optical depth at $E$ = 0.2 TeV is not shown; it is at the level of 5\% at 1 TeV and decreases with wavelength.  The mid-IR limits on the EBL are monotonously decreasing with wavelength and above the limits inferred from fluctuation-analysis of the EBL by \cite{CIBHauserDwek01}. The upper limits in the optical to near-IR  are in conflict with the detections claimed by \citet{IRTS_Matsumoto2000} and also with the HHH scenario by \citet{EBL_Dwek_Krennrich04}. An increase in the upper limits in this region would be achieved if the intrinsic source spectrum peaked in the 0.7-2 TeV region or if the optical depth at 0.2 \mm\ is substantially higher. 

\cite{IRTS_Matsumoto2000} argues that his measurements are inconsistent with galaxy evolution models and much higher than what can be accounted for by the observation of galaxy populations. Already at 2.2 \mm\ the flux is higher than that claimed by \citep{DIRBE-2MASS_Wright01} from COBE and 2MASS data. They found that zodiacal emission modeling as their main uncertainty and responsible for the large fluxes. 

The limits on the EBL derived here are compared with others, derived also with VHE gamma-ray observations, in Fig.~\ref{fig:compare_upper_limits}. Other upper limits are generally lower; this is due to the assumption of a continuous EBL spectrum. In some cases, the upper limits are in conflict with measurements using optical methods.
%
%
%
%
\section{Conclusion\label{sec:conclusion}}
The EBL is difficult to measure directly because of bright foreground radiation originating within the Solar System. VHE gamma-rays are very penetrating and shine through most of the dust in the Galaxy and the Solar system. On cosmological distances, pair production with optical and infrared radiation attenuates the gamma-ray flux. Therefore, if the VHE source spectrum is known, the EBL density can be inferred from the measured attenuation in the spectra. Unfortunately, the known extragalactic sources of VHE gamma rays are highly variable and modeling of the source spectrum is difficult. To estimate the intrinsic source spectrum, the strongest flare spectra of the six sources were compared. These were found to be similar once absorption by the EBL was accounted for. Therefore, the zeroth order assumption was made that the VHE flare spectra of BL Lacs are actually the same; with differences in attenuation arising solely due to the different distances to the objects. This simple, and almost experimental, way to infer the intrinsic spectrum at the chosen energy of 0.2 TeV, results in a spectral index expected from the physics of TeV blazars, see for example \citep{Predicted_Stecker96, modeling_Krawczynski02, Modeling_Konopelko03}.

The method developed here to derive upper limits on the EBL does not hinge on an assumed EBL spectral shape. In detail, it consists of the following three steps: 
(1) The optical depth for gamma-rays was derived at discrete spectral energy points from the difference between the measured gamma-ray spectrum and the assumed intrinsic spectrum. With better understanding of the intrinsic VHE blazar spectra, the optical depth, and hence the resulting upper limits, can be determined more accurately.
(2) The density of EBL photon with a single wavelength was found that reproduces the measured optical depth at a given gamma-ray energy. The EBL wavelength was fixed so that the pair production is most efficient, requiring the smallest density of EBL photons.
(3) The derived monochromatic EBL density represents an upper limit to the true EBL density. The true EBL spectrum is extended and absorption of a gamma ray is produced by EBL photons over a range of wavelengths, thereby requiring a lower density than a monoenergetic EBL to produce the same optical depth. 

\begin{figure}
\plottwo{f4a.eps}{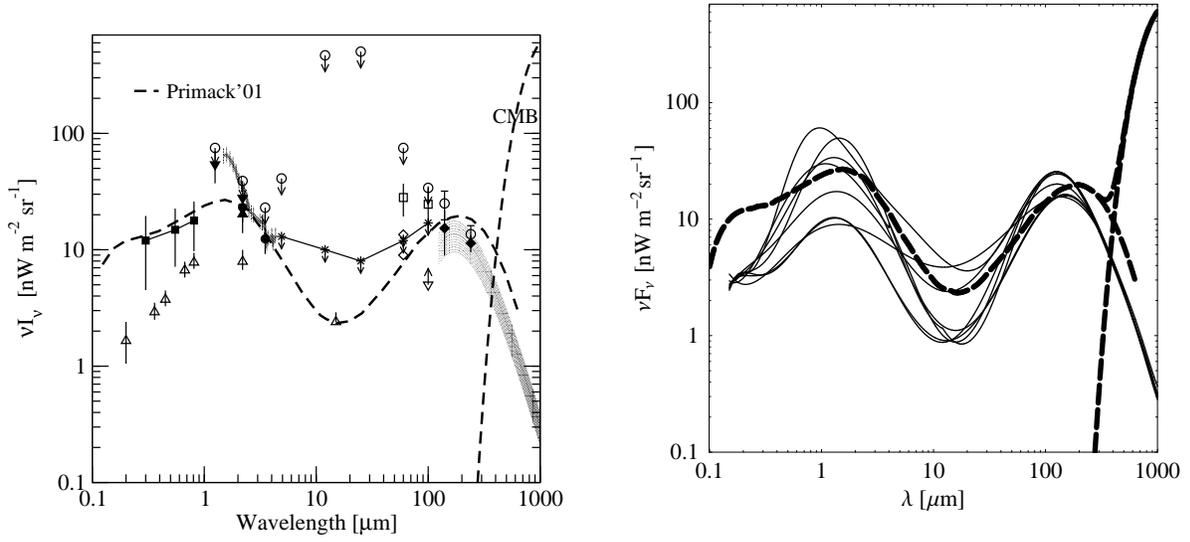}
\caption{The Primack model with CMB flux added (\emph{dashed line}). \emph{Left:} with measurements and upper limits as shown in Fig.\ref{fig:ebl_ir} (\emph{left graph}). \emph{Right:} with the allowed EBL scenarios given in \citep{EBL_Dwek_Krennrich04} (\emph{thin lines}).}
 \label{fig:ebl_models}
\end{figure}

\begin{figure}
\epsscale{.7}
\plotone{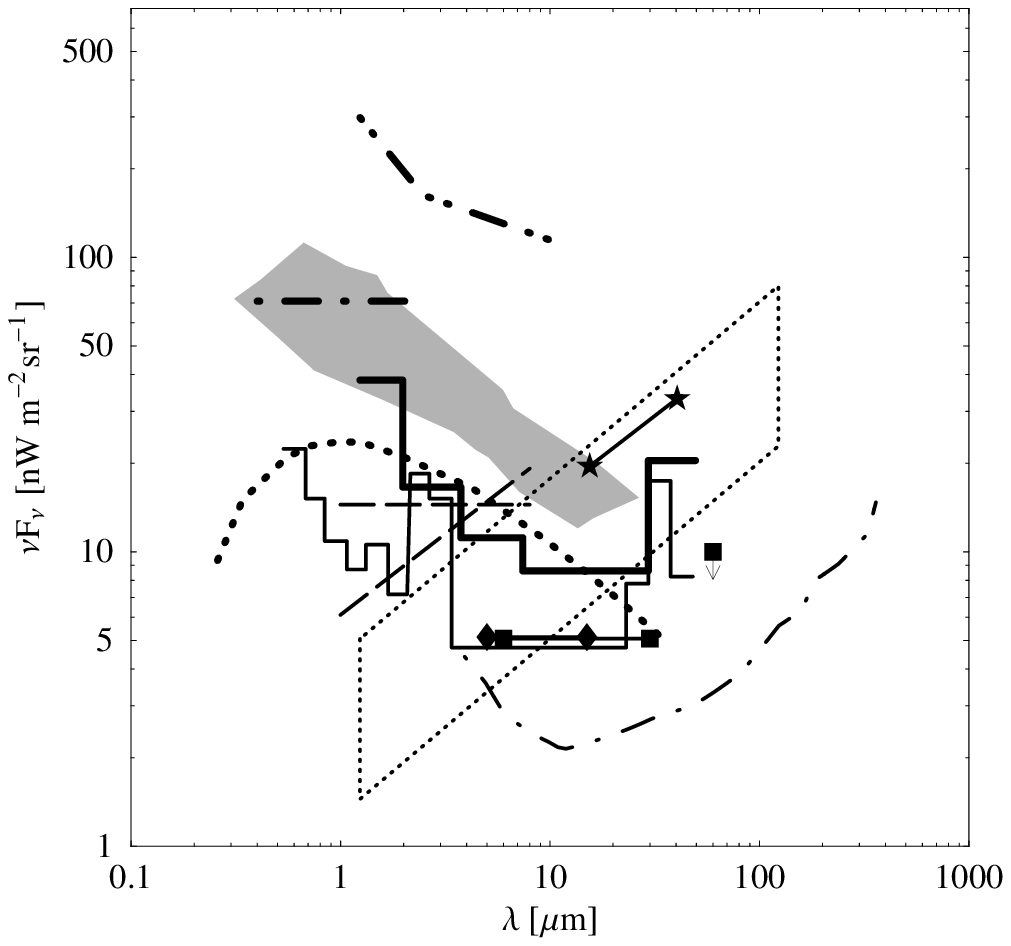}
  \caption{The \emph{gray region} shows the 98\% confidence level upper limits of the EBL flux derived in this work. Upper limits from other publications are as follows: \citet{upper_limits_Stecker93} (\emph{dash}), \citet{IR_determination_Dwek94} (\emph{line with stars}), \citet{Upper_limit_Biller95} (\emph{thick line: dot-dot-dash}), \citet{Mrk501_EBL_Funk98} (\emph{thin line: dot-dot-dash}), \citet{EBL-Mrk501_Stanev98} (\emph{thin line}), \citet{EBL_limits_Biller98} (\emph{thick line}), \citet{Constraints_Guy2000} (\emph{dot-dash}), \citet{EBL_Vassiliev00} (\emph{thick dotted line}), \citet{Dust_Dwek00} (\emph{line and single point with squares}, \citet{EBL-Mrk501_Renault01} (\emph{line with diamonds}) . 
The (\emph{dotted parallelogram}) shows the EBL flux estimate of \citet{Estimate_deJager94}. All flux values have been rescaled to $H_0=71$ km/s/Mpc. }
 \label{fig:compare_upper_limits}
\end{figure}

The optical depths measured from the six blazars fall in line with predictions from a range of EBL models. In particular, for Mrk~421 the measured optical depths represent about the average of the allowed EBL scenarios given by \citet{EBL_Dwek_Krennrich04}. The upper limits derived here are substantially above other limits that were derived assuming a continuous (or stepwise continuous) EBL, see Fig.~\ref{fig:compare_upper_limits}. However, this does not mean that the limits derived here are weak; to properly compare them with EBL models, upper limits must be calculated from the EBL models first. This was done for one EBL model, shown in Fig.~\ref{fig:ebl_models} to represent an average of the allowed EBL scenarios, and the limits derived from the model and the measurements are in agreement.

The constraining power of the upper limits depends on the EBL spectrum: if the spectrum is steeply falling with wavelength, the upper limits are more conservative at shorter wavelength than if the EBL spectrum is flat. The unambiguous and statistically correct determination of 98\% CI upper limits from combining the results of the six blazars is difficult because an average intrinsic blazar spectrum was used, derived from the observed spectral steepening with redshift. The upper limits derived here should be considered as a whole and not from one blazar alone.

The derived upper limits question one particular set of EBL measurements \citep{IRTS_Matsumoto2000} that are very high in the near infrared waveband and which galaxy formation models are typically not able to reproduce either. 

In principle, one could extend the method described here to derive more constraining upper limits on the EBL density by reiterating a process of calculating upper limits and then using the upper-limit spectrum as an input to a second iteration of calculating upper limits. To do better and measure the EBL density, requires accurate knowledge of the intrinsic and the absorbed source spectra over at least two decades of energy. Reduced error bars on the measured spectrum will be achieved with the new generation of telescopes coming online now. However, the mechanism for production of VHE gamma rays, and hence the intrinsic source spectrum, is still under considerable debate. Unless a theoretical model is accepted or a ``standard candle'' at cosmological distances is identified with a well understood gamma-ray spectrum, the spectra from all these sources will only provide upper limits on the EBL flux. Only through further observation at all wavelengths will the gamma-ray production mechanism be understood. The GLAST satellite, to be launched in 2006, will cover the energy range from 0.1-100 GeV. Together with the low threshold of about 100 GeV for the new arrays of imaging Cherenkov telescope such as HESS, VERITAS, and CANGAROO, simultaneous flux measurements can be performed over almost six decades in energy.



\acknowledgments
The author acknowledges the award of a Smithsonian Predoctoral Fellowship and the support of the U. S. Department of Energy. Also, I thank T. C. Weekes for helpful discussions and the anonymous referee for valuable comments.

\clearpage



%

\clearpage
\begin{figure}
\plotone{f2.eps}
\caption{The very high energy flare spectra of six AGN. The \emph{shaded region} shows the fit and 68\% CI for 2 independent parameters. Mrk~421 from \cite{spect_var_Mrk421_Krennrich02} (\emph{open diamonds}), Mrk~501 from \cite{Mrk501_Samuelson98} (\emph{filled diamonds}), 1ES~2344+514 (\emph{filled circles}), 1ES~1959+650 (\emph{open circles}), PKS~2155-413 in 2003 (\emph{left triangles}) \citep{HESS_astroph_2155},  H~1426+428 from \cite{spectrum_H1426_Petry02} (\emph{filled squares}) and \cite{HEGRA_H1426_Aharonian03} (\emph{open squares}). For clarity, the flux level of 1ES~2344, 1ES~1959, and H~1426 has been reduced by a factor of 2, 10, and 10, respectively.}
 \label{fig:spectra_AGN}
\end{figure}

\clearpage

\clearpage

\clearpage


\clearpage

\clearpage

\clearpage

\clearpage

\clearpage

\clearpage

\clearpage

\clearpage

\clearpage


\clearpage








\begin{thebibliography}{70}
\expandafter\ifx\csname natexlab\endcsname\relax\def\natexlab#1{#1}\fi
\expandafter\ifx\csname url\endcsname\relax
  \def\url#1{{\tt #1}}\fi

\bibitem[{Aharonian} et~al.(2002{\natexlab{a}}){Aharonian}, {Akhperjanian},
  {Barrio}, {Beilicke}, {Bernl{\" o}hr}, {B{\" o}rst}, {Bojahr}, {Bolz},
  {Contreras}, {Cornils}, {Cortina}, {Denninghoff}, {Fonseca}, {Girma},
  {Gonzalez}, {G{\" o}tting}, {Heinzelmann}, et~al.]{HEGRA_H1426_Aharonian02}
{Aharonian}~F, {Akhperjanian}~A, {Barrio}~J, et~al.
\newblock {TeV gamma rays from the blazar H 1426+428 and the diffuse
  extragalactic background radiation}.
\newblock {\em \aap}, 384:\penalty0 L23--L26, Mar. 2002{\natexlab{a}}.

\bibitem[{Aharonian} et~al.(2002{\natexlab{b}}){Aharonian}, {Akhperjanian},
  {Beilicke}, {Bernl{\" o}hr}, {B{\" o}rst}, {Bojahr}, {Bolz}, {Coarasa},
  {Contreras}, {Cortina}, {Costamante}, {Denninghoff}, {Fonseca}, {Girma},
  {G{\" o}tting}, {Heinzelmann}, {Hermann}, {Heusler}, {Hofmann}, {Horns},
  {Wiedner}, {Wittek}, and {Remillard}]{spect_var_Mrk421_Aharonian02}
{Aharonian}~F, {Akhperjanian}~A, {Beilicke}~M, et~al.
\newblock {Variations of the TeV energy spectrum at different flux levels of
  Mkn 421 observed with the HEGRA system of Cherenkov telescopes}.
\newblock {\em \aap}, 393:\penalty0 89--99, Oct. 2002{\natexlab{b}}.

\bibitem[{Aharonian} et~al.(2003{\natexlab{a}}){Aharonian}, {Akhperjanian},
  {Beilicke}, {Bernl{\" o}hr}, {B{\" o}rst}, {Bojahr}, {Bolz}, {Coarasa},
  {Contreras}, {Cortina}, {Costamante}, {Denninghoff}, {Fonseca},
  et~al.]{HEGRA_H1426_Aharonian03}
{Aharonian}~F, {Akhperjanian}~A, {Beilicke}~M, et~al.
\newblock {Observations of H1426+428 with HEGRA. Observations in 2002 and
  reanalysis of 1999\&2000 data}.
\newblock {\em \aap}, 403:\penalty0 523--528, May 2003{\natexlab{a}}.

\bibitem[{Aharonian} et~al.(2003{\natexlab{b}}){Aharonian}, {Akhperjanian},
  {Beilicke}, {Bernl{\" o}hr}, {B{\" o}rst}, {Bojahr}, {Bolz}, {Coarasa},
  {Contreras}, {Cortina}, {Denninghoff}, {Fonseca}, {Girma}, {G{\" o}tting},
  {Heinzelmann}, {Hermann}, {Heusler}, {Hofmann}, et~al.]{1es1959Aharonian03}
{Aharonian}~F, {Akhperjanian}~A, {Beilicke}~M, et~al.
\newblock {Detection of TeV gamma-rays from the BL Lac 1ES 1959+650 in its low
  states and during a major outburst in 2002}.
\newblock {\em \aap}, 406:\penalty0 L9--L13, July 2003{\natexlab{b}}.

\bibitem[{Aharonian} et~al.(2004)]{HESS_astroph_2155}
{Aharonian}~F et~al.
\newblock {H.E.S.S. Observations of PKS 2155-304}.
\newblock {\em ArXiv Astrophysics e-prints}, Nov. 2004.
\newblock astro-ph/0411582.

\bibitem[{Aharonian} et~al.(1999){Aharonian}, {Akhperjanian}, {Barrio},
  {Bernl{\" o}hr}, {Bojahr}, {Calle}, {Contreras}, {Cortina}, {Daum},
  {Deckers}, {Denninghoff}, {Fonseca}, {Gonzalez}, {Heinzelmann}, {Wiedner},
  {Willmer}, and {Wittek}]{time_averaged_Mrk501_Aharonian99}
{Aharonian}~FA, {Akhperjanian}~AG, {Barrio}~JA, et~al.
\newblock {The time averaged TeV energy spectrum of MKN 501 of the
  extraordinary 1997 outburst as measured with the stereoscopic Cherenkov
  telescope system of HEGRA}.
\newblock {\em \aap}, 349:\penalty0 11--28, Sept. 1999.

\bibitem[{Armand} et~al.(1994){Armand}, {Milliard}, and
  {Deharveng}]{FarUV_EBL_Armand_94}
{Armand}~C, {Milliard}~B, {Deharveng}~JM.
\newblock {Far-ultraviolet diffuse background and galaxy counts}.
\newblock {\em \aap}, 284:\penalty0 12--16, Apr. 1994.

\bibitem[{Bennett} et~al.(2003){Bennett}, {Halpern}, {Hinshaw}, {Jarosik},
  {Kogut}, {Limon}, {Meyer}, {Page}, {Spergel}, {Tucker}, {Wollack}, {Wright},
  {Barnes}, {Greason}, {Hill}, {Komatsu}, {Nolta}, {Odegard}, {Peiris},
  {Verde}, and {Weiland}]{WMAPBennett03}
{Bennett}~CL, {Halpern}~M, {Hinshaw}~G, et~al.
\newblock {First-Year Wilkinson Microwave Anisotropy Probe (WMAP) Observations:
  Preliminary Maps and Basic Results}.
\newblock {\em \apjs}, 148:\penalty0 1--27, Sept. 2003.

\bibitem[{Bernstein} et~al.(2002){Bernstein}, {Freedman}, and
  {Madore}]{HST_Campanas_1_Bernstein02}
{Bernstein}~RA, {Freedman}~WL, {Madore}~BF.
\newblock {The First Detections of the Extragalactic Background Light at 3000,
  5500, and 8000 {\AA}. I. Results}.
\newblock {\em \apj}, 571:\penalty0 56--84, May 2002.

\bibitem[{Biller}(1995)]{modified_AGN_spectra_Biller95}
{Biller}~SD.
\newblock {Implications of modified AGN spectra due to absorption by infrared
  photons}.
\newblock {\em \app}, 3:\penalty0 385--403, Aug. 1995.

\bibitem[{Biller} et~al.(1995){Biller}, {Akerlof}, {Buckley}, {Cawley},
  {Chantell}, {Fegan}, {Fennell}, {Gaidos}, {Hillas}, {Kerrick}, {Lamb},
  {Lewis}, {Meyer}, {Mohanty}, {O'Flaherty}, {Punch}, {Reynolds}, {Rose},
  {Rovero}, {Schubnell}, {Sembroski}, {Weekes}, and
  {Wilson}]{Upper_limit_Biller95}
{Biller}~SD, {Akerlof}~CW, {Buckley}~J, et~al.
\newblock {An upper limit to the infrared background from observations of TeV
  gamma rays}.
\newblock {\em \apj}, 445:\penalty0 227--230, May 1995.

\bibitem[{Biller} et~al.(1998){Biller}, {Buckley}, {Burdett}, {Bussons Gordo},
  {Carter-Lewis}, {Fegan}, {Finley}, {Gaidos}, {Hillas}, {Krennrich}, {Lamb},
  {Lessard}, {McEnery}, {Mohanty}, {Quinn}, {Rodgers}, {Rose}, {Samuelson},
  {Sembroski}, {Skelton}, {Weekes}, and {Zweerink}]{EBL_limits_Biller98}
{Biller}~SD, {Buckley}~J, {Burdett}~A, et~al.
\newblock {New Limits to the Infrared Background: Bounds on Radiative Neutrino
  Decay and on Contributions of Very Massive Objects to the Dark Matter
  Problem}.
\newblock {\em Physical Review Letters}, 80:\penalty0 2992--2995, Apr. 1998.

\bibitem[{Blain} and {Phillips}(2002)]{60mu_EBL_AGN_Blain02}
{Blain}~AW, {Phillips}~TG.
\newblock {The 60-{$\mu$}m extragalactic background radiation intensity,
  dust-enshrouded active galactic nuclei and the assembly of groups and
  clusters of galaxies}.
\newblock {\em \mnras}, 333:\penalty0 222--230, June 2002.

\bibitem[{Bradbury} et~al.(1997){Bradbury}, {Deckers}, {Petry}, {Konopelko},
  {Aharonian}, {Akhperjanian}, {Barrio}, {Beglarian}, {Beteta}, {Contreras},
  and {Wirth}]{Mrk501_Bradbury97}
{Bradbury}~SM, {Deckers}~T, {Petry}~D, et~al.
\newblock {Detection of {$\gamma$}-rays above 1.5TeV from MKN 501.}
\newblock {\em \aap}, 320:\penalty0 L5--L8, Apr. 1997.

\bibitem[{Breit} and {Wheeler}(1934)]{PairBreit34}
{Breit}~G, {Wheeler}~JA.
\newblock {Collision of Two Light Quanta}.
\newblock {\em Physical Review}, 46:\penalty0 1087, 1934.

\bibitem[{Cambr{\' e}sy} et~al.(2001){Cambr{\' e}sy}, {Reach}, {Beichman}, and
  {Jarrett}]{CIB_at_1.25_2.2_Cambresky01}
{Cambr{\' e}sy}~L, {Reach}~WT, {Beichman}~CA, et~al.
\newblock {The Cosmic Infrared Background at 1.25 and 2.2 Microns Using DIRBE
  and 2MASS: A Contribution Not Due to Galaxies?}
\newblock {\em \apj}, 555:\penalty0 563--571, July 2001.

\bibitem[{Catanese} et~al.(1998){Catanese}, {Akerlof}, {Badran}, {Biller},
  {Bond}, {Boyle}, {Bradbury}, {Buckley}, {Burdett}, {Bussons Gordo},
  {Carter-Lewis}, {Cawley}, {Connaughton}, {Fegan}, {Finley}, {Gaidos}, {Hall},
  {Hillas}, {Krennrich}, {Lamb}, {Lessard}, {Masterson}, {McEnery}, {Mohanty},
  {Quinn}, {Rodgers}, {Rose}, {Samuelson}, {Schubnell}, {Sembroski},
  {Srinivasan}, {Weekes}, {Wilson}, and {Zweerink}]{2344Catanese98}
{Catanese}~M, {Akerlof}~CW, {Badran}~HM, et~al.
\newblock {Discovery of Gamma-Ray Emission above 350 GeV from the BL Lacertae
  Object 1ES 2344+514}.
\newblock {\em \apj}, 501:\penalty0 616, July 1998.

\bibitem[{Chadwick} et~al.(1999){Chadwick}, {Lyons}, {McComb}, {Orford},
  {Osborne}, {Rayner}, {Shaw}, {Turver}, and {Wieczorek}]{pks2155_chadwick99}
{Chadwick}~PM, {Lyons}~K, {McComb}~TJL, et~al.
\newblock {Very High Energy Gamma Rays from PKS 2155-304}.
\newblock {\em \apj}, 513:\penalty0 161--167, Mar. 1999.

\bibitem[{Costamante} et~al.(2001){Costamante}, {Ghisellini}, {Giommi},
  {Tagliaferri}, {Celotti}, {Chiaberge}, {Fossati}, {Maraschi}, {Tavecchio},
  {Treves}, and {Wolter}]{stretching_Costamante01}
{Costamante}~L, {Ghisellini}~G, {Giommi}~P, et~al.
\newblock {Extreme synchrotron BL Lac objects. Stretching the blazar sequence}.
\newblock {\em \aap}, 371:\penalty0 512--526, May 2001.

\bibitem[{de Jager} et~al.(1994){de Jager}, {Stecker}, and
  {Salamon}]{Estimate_deJager94}
{de Jager}~OC, {Stecker}~FW, {Salamon}~MH.
\newblock {Estimate of the Intergalactic Infrared Radiation Field from
  Gamma-Ray Observations of the Galaxy MARKARIAN:421}.
\newblock {\em \nat}, 369:\penalty0 294--296, May 1994.

\bibitem[{Dwek}(2001)]{Dust_Dwek00}
{Dwek}~E.
\newblock {The Role of Dust in Producing the Extragalactic Diffuse Background}.
\newblock In {\em IAU Symposium}, pages 389+, Jan. 2001.

\bibitem[{Dwek} et~al.(1998){Dwek}, {Arendt}, {Hauser}, {Fixsen}, {Kelsall},
  {Leisawitz}, {Pei}, {Wright}, {Mather}, {Moseley}, {Odegard}, {Shafer},
  {Silverberg}, and {Weiland}]{COBE_DIRBE_cosmological_implications_Dwek98}
{Dwek}~E, {Arendt}~RG, {Hauser}~MG, et~al.
\newblock {The COBE Diffuse Infrared Background Experiment Search for the
  Cosmic Infrared Background. IV. Cosmological Implications}.
\newblock {\em \apj}, 508:\penalty0 106--122, Nov. 1998.

\bibitem[{Dwek} and {Krennrich}(2005)]{EBL_Dwek_Krennrich04}
{Dwek}~E, {Krennrich}~F.
\newblock {Simultaneous constraints on the spectrum of the extragalactic
  background light and the intrinsic TeV spectra of Mrk 421, Mrk 501, and
  H1426+428}.
\newblock {\em \apj}, 618:\penalty0 657, 2005.

\bibitem[{Dwek} and {Slavin}(1994)]{IR_determination_Dwek94}
{Dwek}~E, {Slavin}~J.
\newblock {On the determination of the cosmic infrared background radiation
  from the high-energy spectrum of extragalactic gamma-ray sources}.
\newblock {\em \apj}, 436:\penalty0 696--704, Dec. 1994.

\bibitem[{Elbaz} et~al.(2002){Elbaz}, {Cesarsky}, {Chanial}, {Aussel},
  {Franceschini}, {Fadda}, and {Chary}]{CIB-ISOCAM_Elbaz02}
{Elbaz}~D, {Cesarsky}~CJ, {Chanial}~P, et~al.
\newblock {The bulk of the cosmic infrared background resolved by ISOCAM}.
\newblock {\em \aap}, 384:\penalty0 848--865, Mar. 2002.

\bibitem[{Falcone}(2004)]{H1426_Falcone04}
{Falcone}~AD.
\newblock {X-ray and TeV observations of the extreme BL Lac AGN H1426+428
  during 2002}.
\newblock {\em New Astronomy Review}, 48:\penalty0 415--417, Apr. 2004.

\bibitem[{Finkbeiner} et~al.(2000){Finkbeiner}, {Davis}, and
  {Schlegel}]{IRB_60_100_with_DIRBE_Finkbeiner2000}
{Finkbeiner}~DP, {Davis}~M, {Schlegel}~DJ.
\newblock {Detection of a Far-Infrared Excess with DIRBE at 60 and 100
  Microns}.
\newblock {\em \apj}, 544:\penalty0 81--97, Nov. 2000.

\bibitem[{Fixsen} et~al.(1998){Fixsen}, {Dwek}, {Mather}, {Bennett}, and
  {Shafer}]{FIRAS_spectrum_Fixsen98}
{Fixsen}~DJ, {Dwek}~E, {Mather}~JC, et~al.
\newblock {The Spectrum of the Extragalactic Far-Infrared Background from the
  COBE FIRAS Observations}.
\newblock {\em \apj}, 508:\penalty0 123--128, Nov. 1998.

\bibitem[{Franceschini} et~al.(1991){Franceschini}, {Toffolatti}, {Mazzei},
  {Danese}, and {de Zotti}]{galaxy_counts_Franceschini91}
{Franceschini}~A, {Toffolatti}~L, {Mazzei}~P, et~al.
\newblock {Galaxy counts and contributions to the background radiation from 1
  micron to 1000 microns}.
\newblock {\em \aaps}, 89:\penalty0 285--310, Aug. 1991.

\bibitem[{Funk} et~al.(1998){Funk}, {Magnussen}, {Meyer}, {Rhode},
  {Westerhoff}, and {Wiebel-Sooth}]{Mrk501_EBL_Funk98}
{Funk}~B, {Magnussen}~N, {Meyer}~H, et~al.
\newblock {An upper limit on the infrared background density from HEGRA data on
  MKN 501}.
\newblock {\em Astroparticle Physics}, 9:\penalty0 97--103, June 1998.

\bibitem[{Guy} et~al.(2000){Guy}, {Renault}, {Aharonian}, {Rivoal}, and
  {Tavernet}]{Constraints_Guy2000}
{Guy}~J, {Renault}~C, {Aharonian}~FA, et~al.
\newblock {Constraints on the cosmic infra-red background based on BeppoSAX and
  CAT spectra of Markarian 501}.
\newblock {\em \aap}, 359:\penalty0 419--428, July 2000.

\bibitem[{Hauser} et~al.(1998){Hauser}, {Arendt}, {Kelsall}, {Dwek}, {Odegard},
  {Weiland}, {Freudenreich}, {Reach}, {Silverberg}, {Moseley}, {Pei}, {Lubin},
  {Mather}, {Shafer}, {Smoot}, {Weiss}, {Wilkinson}, and
  {Wright}]{COBE_DIRBE_Hauser98}
{Hauser}~MG, {Arendt}~RG, {Kelsall}~T, et~al.
\newblock {The COBE Diffuse Infrared Background Experiment Search for the
  Cosmic Infrared Background. I. Limits and Detections}.
\newblock {\em \apj}, 508:\penalty0 25--43, Nov. 1998.

\bibitem[{Hauser} and {Dwek}(2001)]{CIBHauserDwek01}
{Hauser}~MG, {Dwek}~E.
\newblock {The Cosmic Infrared Background: Measurements and Implications}.
\newblock {\em \araa}, 39:\penalty0 249--307, 2001.

\bibitem[{Holder} et~al.(2003){Holder}, {Bond}, {Boyle}, {Bradbury}, {Buckley},
  {Carter-Lewis}, {Cui}, {Dowdall}, {Duke}, {de la Calle Perez}, {Falcone},
  {Fegan}, {Fegan}, {Finley}, {Fortson}, {Gaidos}, {Gibbs}, {Gammell}, {Hall},
  {Hall}, {Hillas}, {Horan}, {Jordan}, {Kertzman}, {Kieda}, {Kildea}, {Knapp},
  {Kosack}, {Krawczynski}, {Krennrich}, {LeBohec}, {Linton}, {Lloyd-Evans},
  {Moriarty}, {M{\" u}ller}, {Nagai}, {Ong}, {Page}, {Pallassini}, {Petry},
  {Power-Mooney}, {Quinn}, {Rebillot}, {Reynolds}, {Rose}, {Schroedter},
  {Sembroski}, {Swordy}, {Vassiliev}, {Wakely}, {Walker}, and
  {Weekes}]{1es1959Holder03}
{Holder}~J, {Bond}~IH, {Boyle}~PJ, et~al.
\newblock {Detection of TeV Gamma Rays from the BL Lacertae Object 1ES 1959+650
  with the Whipple 10 Meter Telescope}.
\newblock {\em \apjl}, 583:\penalty0 L9--L12, Jan. 2003.

\bibitem[{Horan} et~al.(2002){Horan}, {Badran}, {Bond}, {Bradbury}, {Buckley},
  {Carson}, {Carter-Lewis}, {Catanese}, {Cui}, {Dunlea}, {Das}, {de la Calle
  Perez}, {D'Vali}, {Fegan}, {Fegan}, {Finley}, {Gaidos}, {Gibbs},
  {Gillanders}, {Hall}, {Hillas}, {Wakely}, and {Weekes}]{H1426_horan02}
{Horan}~D, {Badran}~HM, {Bond}~IH, et~al.
\newblock {Detection of the BL Lacertae Object H1426+428 at TeV Gamma-Ray
  Energies}.
\newblock {\em \apj}, 571:\penalty0 753--762, June 2002.

\bibitem[{Kashlinsky} et~al.(1996){Kashlinsky}, {Mather}, and
  {Odenwald}]{fluctuationCOBE_Kashlinsky96}
{Kashlinsky}~A, {Mather}~JC, {Odenwald}~S.
\newblock {Clustering of the Diffuse Infrared Light from the COBE DIRBE Maps:
  an All-Sky Survey of C(0)}.
\newblock {\em \apjl}, 473:\penalty0 L9+, Dec. 1996.

\bibitem[{Kashlinsky} and {Odenwald}(2000)]{FluctuationCOBE_Kashlinsky2000}
{Kashlinsky}~A, {Odenwald}~S.
\newblock {Clustering of the Diffuse Infrared Light from the COBE DIRBE Maps.
  III. Power Spectrum Analysis and Excess Isotropic Component of Fluctuations}.
\newblock {\em \apj}, 528:\penalty0 74--95, Jan. 2000.

\bibitem[{Konopelko} et~al.(2003){Konopelko}, {Mastichiadis}, {Kirk}, {de
  Jager}, and {Stecker}]{Modeling_Konopelko03}
{Konopelko}~A, {Mastichiadis}~A, {Kirk}~J, et~al.
\newblock {Modeling the TeV Gamma-Ray Spectra of Two Low-Redshift Active
  Galactic Nuclei: Markarian 501 and Markarian 421}.
\newblock {\em \apj}, 597:\penalty0 851--859, Nov. 2003.

\bibitem[{Krawczynski} et~al.(2002){Krawczynski}, {Coppi}, and
  {Aharonian}]{modeling_Krawczynski02}
{Krawczynski}~H, {Coppi}~PS, {Aharonian}~F.
\newblock {Time-dependent modelling of the Markarian 501 X-ray and TeV
  gamma-ray data taken during 1997 March and April}.
\newblock {\em \mnras}, 336:\penalty0 721--735, Nov. 2002.

\bibitem[{Krawczynski} et~al.(2004){Krawczynski}, {Hughes}, {Horan},
  {Aharonian}, {Aller}, {Aller}, {Boltwood}, {Buckley}, {Coppi}, {Fossati},
  {G{\" o}tting}, {Holder}, {Horns}, {Kurtanidze}, {Marscher}, {Nikolashvili},
  {Remillard}, {Sadun}, and {Schr{\" o}der}]{multi_wave_1es1959_Krawczynski04}
{Krawczynski}~H, {Hughes}~SB, {Horan}~D, et~al.
\newblock {Multiwavelength Observations of Strong Flares from the TeV Blazar
  1ES 1959+650}.
\newblock {\em \apj}, 601:\penalty0 151--164, Jan. 2004.

\bibitem[{Krennrich} et~al.(2001){Krennrich}, {Badran}, {Bond}, {Bradbury},
  {Buckley}, {Carter-Lewis}, {Catanese}, {Cui}, {Dunlea}, {Das}, {de la Calle
  Perez}, {Fegan}, {Fegan}, {Finley}, {Gaidos}, {Gibbs}, {Gillanders}, {Hall},
  {Hillas}, {Holder}, {Horan}, {Jordan}, {Kertzman}, {Kieda}, {Kildea},
  {Knapp}, {Kosack}, {Lang}, {LeBohec}, {McKernan}, {Moriarty}, {M{\" u}ller},
  {Ong}, {Pallassini}, {Petry}, {Quinn}, {Reay}, {Reynolds}, {Rose},
  {Sembroski}, {Sidwell}, {Stanton}, {Swordy}, {Vassiliev}, {Wakely}, and
  {Weekes}]{Mrk421_Krennrich01}
{Krennrich}~F, {Badran}~HM, {Bond}~IH, et~al.
\newblock {Cutoff in the TeV Energy Spectrum of Markarian 421 during Strong
  Flares in 2001}.
\newblock {\em \apjl}, 560:\penalty0 L45--L48, Oct. 2001.

\bibitem[{Krennrich} et~al.(2002){Krennrich}, {Bond}, {Bradbury}, {Buckley},
  {Carter-Lewis}, {Cui}, {de la Calle Perez}, {Fegan}, {Fegan}, {Finley},
  {Gaidos}, {Gibbs}, {Gillanders}, {Hall}, {Hillas}, {Holder}, {Horan},
  {Jordan}, {Kertzman}, {Kieda}, {Kildea}, {Knapp}, {Kosack}, {Lang},
  {LeBohec}, {Moriarty}, {M{\" u}ller}, {Ong}, {Pallassini}, {Petry}, {Quinn},
  {Reay}, {Reynolds}, {Rose}, {Sembroski}, {Sidwell}, {Stanton}, {Swordy},
  {Vassiliev}, {Wakely}, and {Weekes}]{spect_var_Mrk421_Krennrich02}
{Krennrich}~F, {Bond}~IH, {Bradbury}~SM, et~al.
\newblock {Discovery of Spectral Variability of Markarian 421 at TeV Energies}.
\newblock {\em \apjl}, 575:\penalty0 L9--L13, Aug. 2002.

\bibitem[{Krennrich} and {Dwek}(2003)]{EBL_Krennrich03}
{Krennrich}~F, {Dwek}~E.
\newblock {Intrinsic Spectra of the TeV Blazars Mrk 421 and Mrk 501}.
\newblock In {\em 28th \ICRC}, 2003.
\newblock astro-ph/0305418.

\bibitem[{Lagache} et~al.(1999){Lagache}, {Abergel}, {Boulanger}, {D{\'
  e}sert}, and {Puget}]{COBE_FIRAS_Lagache99}
{Lagache}~G, {Abergel}~A, {Boulanger}~F, et~al.
\newblock {First detection of the warm ionised medium dust emission.
  Implication for the cosmic far-infrared background}.
\newblock {\em \aap}, 344:\penalty0 322--332, Apr. 1999.

\bibitem[{Leinert} et~al.(1998){Leinert}, {Bowyer}, {Haikala}, {Hanner},
  {Hauser}, {Levasseur-Regourd}, {Mann}, {Mattila}, {Reach}, {Schlosser},
  {Staude}, {Toller}, {Weiland}, {Weinberg}, and
  {Witt}]{Diffuse_NSB_reference_Leinert98}
{Leinert}~C, {Bowyer}~S, {Haikala}~LK, et~al.
\newblock {The 1997 reference of diffuse night sky brightness}.
\newblock {\em \aaps}, 127:\penalty0 1--99, Jan. 1998.

\bibitem[{Malkan} and {Stecker}(2001)]{Malkan2001_EmpiricalModel}
{Malkan}~MA, {Stecker}~FW.
\newblock {An Empirically Based Model for Predicting Infrared Luminosity
  Functions, Deep Infrared Galaxy Counts, and the Diffuse Infrared Background}.
\newblock {\em \apj}, 555:\penalty0 641--649, July 2001.

\bibitem[{Matsumoto}(2000)]{IRTS_Matsumoto2000}
{Matsumoto}~T.
\newblock {Observations of Infrared Extragalactic Background Light}.
\newblock {\em The Institute of Space and Astronautical Science Report SP
  No.~14, p.~179-186.}, 14:\penalty0 179--186, Dec. 2000.

\bibitem[{Miville-Desch{\^ e}nes} et~al.(2002){Miville-Desch{\^ e}nes},
  {Lagache}, and {Puget}]{60mu_Miville02}
{Miville-Desch{\^ e}nes}~MA, {Lagache}~G, {Puget}~JL.
\newblock {Power spectrum of the cosmic infrared background at 60 and 100 \mm\
  with IRAS}.
\newblock {\em \aap}, 393:\penalty0 749--756, Oct. 2002.

\bibitem[{Nikishov}(1962)]{AbsorptionNikishov62}
{Nikishov}~AI.
\newblock {Absorption of High-Energy Photons in the Universe}.
\newblock {\em Soviet Physics Jetp}, 14:\penalty0 393--394, Feb. 1962.

\bibitem[{Nishiyama} et~al.(1999)]{1es1959Nishiyama99}
{Nishiyama}~T et~al.
\newblock {1ES1959+650}.
\newblock In {\em 26th International Cosmic Ray Conference}, volume~3, pages
  370--373, 1999.

\bibitem[{Petry} et~al.(2002){Petry}, {Bond}, {Bradbury}, {Buckley},
  {Carter-Lewis}, {Cui}, {Duke}, {de la Calle Perez}, {Falcone}, {Fegan},
  {Fegan}, {Finley}, {Gaidos}, {Gibbs}, {Gammell}, {Hall},
  et~al.]{spectrum_H1426_Petry02}
{Petry}~D, {Bond}~IH, {Bradbury}~SM, et~al.
\newblock {The TeV Spectrum of H1426+428}.
\newblock {\em \apj}, 580:\penalty0 104--109, Nov. 2002.

\bibitem[{Petry} et~al.(1996){Petry}, {Bradbury}, {Konopelko}, {Fernandez},
  {Aharonian}, {Akhperjanian}, {Belgarian}, {Beteta}, {Contreras}, {Cortina},
  et~al.]{Mrk421_Petry96}
{Petry}~D, {Bradbury}~SM, {Konopelko}~A, et~al.
\newblock {Detection of VHE {$\gamma$}-rays from MKN 421 with the HEGRA
  Cherenkov Telescopes.}
\newblock {\em \aap}, 311:\penalty0 L13--L16, July 1996.

\bibitem[{Pozzetti} et~al.(1998){Pozzetti}, {Madau}, {Zamorani}, {Ferguson},
  and {Bruzual A.}]{Hubble_deep_field_Pozzetti98}
{Pozzetti}~L, {Madau}~P, {Zamorani}~G, et~al.
\newblock {High-redshift galaxies in the Hubble Deep Field - II. Colours and
  number counts}.
\newblock {\em \mnras}, 298:\penalty0 1133--1144, Aug. 1998.

\bibitem[{Primack} et~al.(2001){Primack}, {Somerville}, {Bullock}, and
  {Devriendt}]{model_Primack}
{Primack}~JR, {Somerville}~RS, {Bullock}~JS, et~al.
\newblock {Probing Galaxy Formation with High-Energy Gamma-Rays}.
\newblock In {\em High Energy Gamma Ray Astronomy}, volume 558, pages 463--478,
  2001.
\newblock Proceedings of the International Symposium Gamma-2000.

\bibitem[{Punch} et~al.(1992){Punch}, {Akerlof}, {Cawley}, {Chantell}, {Fegan},
  {Fennell}, {Gaidos}, {Hagan}, {Hillas}, {Jiang}, {Kerrick}, {Lamb},
  {Lawrence}, {Lewis}, {Meyer}, {Mohanty}, {O'Flaherty}, {Reynolds}, {Rovero},
  {Schubnell}, {Sembroski}, {Weekes}, and {Wilson}]{Mrk421_Punch92}
{Punch}~M, {Akerlof}~CW, {Cawley}~MF, et~al.
\newblock {Detection of TeV photons from the active galaxy Markarian 421}.
\newblock {\em \nat}, 358:\penalty0 477--478, Aug. 1992.

\bibitem[{Quinn} et~al.(1996){Quinn}, {Akerlof}, {Biller}, {Buckley},
  {Carter-Lewis}, {Cawley}, {Catanese}, {Connaughton}, {Fegan}, {Finley},
  {Gaidos}, {Hillas}, et~al.]{Mrk501_Quinn96}
{Quinn}~J, {Akerlof}~CW, {Biller}~S, et~al.
\newblock {Detection of Gamma Rays with E greater than 300 GeV from Markarian
  501}.
\newblock {\em \apjl}, 456:\penalty0 L83--L86, Jan. 1996.

\bibitem[{Renault} et~al.(2001){Renault}, {Barrau}, {Lagache}, and
  {Puget}]{EBL-Mrk501_Renault01}
{Renault}~C, {Barrau}~A, {Lagache}~G, et~al.
\newblock {New constraints on the cosmic mid-infrared background using TeV
  gamma-ray astronomy}.
\newblock {\em \aap}, 371:\penalty0 771--778, May 2001.

\bibitem[{Salamon} and {Stecker}(1998)]{absorption_lowE_Salamon98}
{Salamon}~MH, {Stecker}~FW.
\newblock {Absorption of High-Energy Gamma Rays by Interactions with
  Extragalactic Starlight Photons at High Redshifts and the High-Energy
  Gamma-Ray Background}.
\newblock {\em \apj}, 493:\penalty0 547--554, Jan. 1998.

\bibitem[{Samuelson} et~al.(1998){Samuelson}, {Biller}, {Bond}, {Boyle},
  {Bradbury}, {Breslin}, {Buckley}, {Burdett}, {Buss'ons Gordo},
  {Carter-Lewis}, {Cantanese}, {Cawley}, {Fegan}, {Finley}, {Gaidos}, {Hall},
  {Hillas}, {Krennrich}, {Lamb}, {Lessard}, {McEnery}, {Masterson}, {Quinn},
  {Rodgers}, {Rose}, {Sembroski}, {Srinivasan}, {Vassiliev}, {Weekes}, and
  {Zweerink}]{Mrk501_Samuelson98}
{Samuelson}~FW, {Biller}~SD, {Bond}~IH, et~al.
\newblock {The TeV Spectrum of Markarian 501}.
\newblock {\em \apjl}, 501:\penalty0 L17--L20, July 1998.

\bibitem[{Schroedter}(2004)]{thesis_Schroedter04}
{Schroedter}~M.
\newblock {\em {The Very High Energy Gamma-Ray Spectra of AGN}}.
\newblock PhD thesis, University of Arizona, 2004.

\bibitem[{Schroedter} et~al.(2005)]{2344_Schroedter05}
{Schroedter}~M et~al.
\newblock {}.
\newblock 2005.
\newblock {to be published}.

\bibitem[{Stanev} and {Franceschini}(1998)]{EBL-Mrk501_Stanev98}
{Stanev}~T, {Franceschini}~A.
\newblock {Constraints on the Extragalactic Infrared Background from Gamma-Ray
  Observations of MRK 501}.
\newblock {\em \apjl}, 494:\penalty0 L159--L162, Feb. 1998.

\bibitem[{Stecker}(1999)]{Steepening_Stecker99}
{Stecker}~FW.
\newblock {Intergalactic extinction of high energy gamma-rays}.
\newblock {\em Astroparticle Physics}, 11:\penalty0 83--91, June 1999.

\bibitem[{Stecker} and {de Jager}(1993)]{upper_limits_Stecker93}
{Stecker}~FW, {de Jager}~OC.
\newblock {New Upper Limits on Intergalactic Infrared Radiation from
  High-Energy Astrophysics}.
\newblock {\em \apjl}, 415:\penalty0 L71--L73, Oct. 1993.

\bibitem[{Stecker} et~al.(1992){Stecker}, {de Jager}, and
  {Salamon}]{TeV_EBL_probe_Stecker92}
{Stecker}~FW, {de Jager}~OC, {Salamon}~MH.
\newblock {TeV gamma rays from 3C 279 - A possible probe of origin and
  intergalactic infrared radiation fields}.
\newblock {\em \apjl}, 390:\penalty0 L49--L52, May 1992.

\bibitem[{Stecker} et~al.(1996){Stecker}, {de Jager}, and
  {Salamon}]{Predicted_Stecker96}
{Stecker}~FW, {de Jager}~OC, {Salamon}~MH.
\newblock {Predicted Extragalactic TeV Gamma-Ray Sources}.
\newblock {\em \apjl}, 473:\penalty0 L75--L78, Dec. 1996.

\bibitem[{Tluczykont} et~al.(2003){Tluczykont}, {G{\"o}tting}, {Heinzelmann},
  and {HEGRA}]{2344Hegra03}
{Tluczykont}~M, {G{\"o}tting}~N, {Heinzelmann}~G, et~al.
\newblock {Observations of 54 Active Galactic Nuclei with the HEGRA Cherenkov
  Telescopes}.
\newblock In {\em International Cosmic Ray Conference}, volume~5, pages
  2547--2550, 2003.

\bibitem[{Vassiliev}(2000)]{EBL_Vassiliev00}
{Vassiliev}~VV.
\newblock {Extragalactic background light absorption signal in the TeV
  gamma-ray spectra of blazars.}
\newblock {\em Astroparticle Physics}, 12:\penalty0 217--238, Jan. 2000.

\bibitem[{Wright}(2001)]{DIRBE-2MASS_Wright01}
{Wright}~EL.
\newblock {DIRBE minus 2MASS: Confirming the Cosmic Infrared Background at 2.2
  Microns}.
\newblock {\em \apj}, 553:\penalty0 538--544, June 2001.

\bibitem[{Wright} and {Reese}(2000)]{COBE_2.2_3.5_Wright2000}
{Wright}~EL, {Reese}~ED.
\newblock {Detection of the Cosmic Infrared Background at 2.2 and 3.5 Microns
  Using DIRBE Observations}.
\newblock {\em \apj}, 545:\penalty0 43--55, Dec. 2000.

\end{thebibliography}
\end{document}